# Chemical inhomogeneities in high-entropy alloys help mitigate the strength-ductility trade-off

Evan Ma* and Chang Liu**

*Center for Alloy Innovation and Design (CAID), State Key Lab for Mechanical Behavior of Materials, Xi'an Jiaotong University, Xi'an, China*

*maen@xjtu.edu.cn

**chang.liu@xjtu.edu.cn

*Metallurgists have long been accustomed to a trade-off between yield strength and tensile ductility. Extending previously known strain-hardening mechanisms, the emerging multi-principal-element alloys (MPEAs) offer additional help in promoting the strength-ductility synergy, towards gigapascal yield strength simultaneously with pure-metal-like tensile ductility. The highly concentrated chemical make-up in these "high-entropy" alloys (HEAs) adds, at ultrafine spatial scale from sub-nanometer to tens of nanometers, inherent chemical inhomogeneities in local composition and local chemical order (LCO). These institute a "nano-cocktail" environment that exerts extra dragging forces, rendering a much wavier motion of dislocation lines (in stick-slip mode) different from dilute solutions. The variable fault energy landscape also makes the dislocation movement sluggish, increasing their chances to hit one another and react to increase entanglement. The accumulation of dislocations (plus faults) dynamically stores obstacles against ensuing dislocation motion to sustain an adequate strain-hardening rate at high flow stresses, delaying plastic instability to enable large (uniform) elongation. The successes summarized advocate MPEAs as an effective recipe towards ultrahigh strength at little expense of tensile ductility. The insight gained also answers the question as to what new mechanical behavior the HEAs have to offer, beyond what has been well documented for traditional metals and solid solutions.*

## 1. Trade-off between yield strength and tensile ductility in metallic materials

Elemental metals and solid solutions based on them boast high tensile ductility (elongation to failure of the order of 50% in uniaxial tension), providing adequate formability for shaping/manufacturing as well as stable flow after yield for ample safety margin against catastrophic fracture. However, the yield strength ($\sigma_y$) of these traditional metals is insufficient for many critical load-bearing structural applications. Several textbook routes are available to raise the $\sigma_y$ up towards gigapascal (GPa) level. Unfortunately, such a gain in strength is often accompanied by a sacrifice in tensile ductility. As a norm, for a metal or alloy strengthened to increase its $\sigma_y$ (e.g., via grain refinement or cold work or alloying), its tensile ductility keeps decreasing with rising strength, from its originally large elongation before strengthening. This is often termed "strength-ductility trade-off", the topic to be covered in this review.

We begin with the light-blue-shaded area [1] in Fig. 1, which covers typical experimental data of $\sigma_y$ versus achievable ductility in a uniaxial tensile test (in terms of both the uniform elongation before strain localization or necking in **a**) and total percent elongation to failure in **b**)), for face-centered cubic (fcc) and body-centered-cubic (bcc) metallic materials at room temperature. A real-world fcc case using Ni processed to various microstructures is shown in Fig. S1 as an example (other metals such as Cu and steels are included in Fig. S2a). The shape of the blue area in Fig. 1 (as well as the banana-shaped ones in Fig. S1) points to a seemingly universal dilemma: one can either reach for $\sigma_y >$ GPa (a benchmark surpassing most commercial steels and nonferrous alloys [1–3]) at some expense of ductility, or relish high tensile ductility while living with the far-from-adequate strength, but having it both ways seems implausible.

The absence of simultaneously high strength and high tensile ductility, i.e., the unoccupied space towards the upper right corner in Fig. 1 and Fig. S1, is widely perceived in the metallurgy field as reflecting an intrinsic conflict between materials strength and toughness. For example, a remark commonly made as a rationale for the trade-off is that the dislocations are hopelessly confused: high strength means major difficulty against their activities, while high ductility demands easy movements of these plasticity carriers. The lack of strength-ductility synergy has then been purported to be perpetual and insurmountable. Such a postulate, while reasonable for brittle elements and intermetallics, is incorrect for many metallic materials including those discussed in this review (such as Ni in Fig. S1 and some of the HEAs). The low tensile ductility should not be misconstrued as equivalent to the loss of dislocation activities altogether like in ceramics and brittle intermetallics, which suffocates plasticity mechanisms and therefore leads to low toughness. In most strengthened metals and solid solutions, dislocations (such as Fig. 1 and the HEAs discussed in this review) continue to function as viable plasticity carriers. An example that they remain widely active is reflected by the sizable elongation in the tensile curves for Ni in Fig. S1. Another case in this regard is that ultra-strong nanocrystalline metals (Ni with 2 GPa strength) can be compression-deformed into pancakes with no fracture, showing a large area under the stress-strain curve [4]. In other words, strengthening to GPa strength does not necessarily forfeit plastic flow capability. In fact, the compressive ductility is not compromised, and it is the tensile ductility that is at stake.

The degradation of tensile ductility at high flow stresses for the metallic materials is a result of an increased propensity for unstable (severely localized) flow in uniaxial tension, specifically from the plastic instability criterion [3] being violated at an earlier stage/strain of plastic deformation. When the $\sigma_y$ is raised, the subsequent tensile flow is stable when, and only when, the strain hardening modulus (i.e., the strain hardening rate $\Theta$, which is the slope of the stress versus strain curve beyond the point of yield) stays adequately high. This is well known from the Hart's criterion [5,6]. Once an instability such as necking in uniaxial tensile deformation (or shear banding, or other forms of strain localization) sets in, severely localized flow wreaks havoc to instigate uncontrollable damage, rather than spreading the strain across the sample volume. Unfortunately, $\Theta$ often fails to keep up proportionally with the $\sigma_y$ increase, for all the textbook mechanism(s) known for strengthening [1,2]. Resolving this dilemma has in fact been a long-sought goal, as well as a stiff challenge, in the pursuit towards optimized/tailored properties via microstructural design.

Nonetheless, the origin of the trade-off, as demystified above for the metallic materials discussed in this review, brings with it a positive implication, because intrinsically the two properties (strength and ductility of metals) are then not mutually exclusive as often purported in the community. We stress again that difficult dislocation motion (high flow stress) should not be portrayed as utterly suppressed dislocation activities. This is fundamentally different from brittle ceramics and intermetallics where the dislocation mechanism is taken away by and large, precluding their ductility and toughness. For metallic materials, as long as the dislocations can be activated by high driving stresses without losing to premature damage, and adequately delocalized to ward off flow localization instability, tenacious dislocation motion can still mediate stable plastic flow to permeate the deforming volume; the resultant high ductility will be amply illustrated using examples in this review.

The possibility to reap the benefits of such an opportunity has been examined in numerous attempts made over the past two decades. Moderate successes improving the balance between strength and tensile ductility have in fact been reported. The first category is for simple metals and less-concentrated alloys, and the strategy applied often invokes heterogeneous nanostructures (in forms of nonuniform distribution or gradient of grain size, or nano-lamella vs micron-scale lateral dimensions) [1], as represented by the pink-shaded area in Fig. 1 (and in Fig. S2). The structural inhomogeneities deployed induce strain gradient inside the material, i.e., the softer regions deform plastically while the harder domains remain elastic. To accommodate this mismatch, high densities of geometrically necessary dislocations (GNDs) are created [1,7], which generate back stresses and impose additional non-local hardening. This non-homogeneous plastic deformation [8] (including strain gradient plasticity) effect, while known for a long time, has recently been re-cast as hetero-deformation induced (HDI) hardening with quantifiable back-stress contributions to the strain hardening [9]. However, while the pink banana-shaped region is shifted up a notch with respect to the original (light-blue) one, even the best cases remained not much above an interpolation line (imagine a straight line) connecting the high strength (but low ductility) and the high ductility (but low strength) end points. What has been achieved is then just another level of trade-off, albeit this time a compromise akin to a rule-of-mixtures average (see Fig. S2a, which is adapted from a recent review [10]). A combination of nanograin-like strength and coarse-grain-like ductility, while hypothesized (Fig. S2b) as an upside-down banana curve bulging towards the upper right corner [11], has remained a target that is yet to be materialized in reality (compare with the experimental findings summarized recently in Fig. S2a). We thus conclude that simple metals, even when tailored structurally heterogeneous, fall short in their prospect to overcome the strength-ductility trade-off, because the structural inhomogeneities are limited in their potential to provide a sufficient strain hardening capacity.

The next strategy to evade the trade-off is by designing complex alloys that incorporate appreciable amounts of solutes and second phases into the host (solvent) metal. Some representative alloys with good strength-ductility combinations are presented in Fig. 1, including solid solution brass (white hollow squares), Ti alloys (violet solid triangles) and high-performance steels (white solid hexagons) including high-nitrogen austenitic ones. While many of these alloys still follow the general trend for the blue and pink regions, some extend into the yellow regime. These latter strong-yet-ductile alloys often benefit from the sequential onset of TWIP, TRIP effects [12,13], and/or formation of high-density dislocation walls, etc., such that multiple strain hardening stages along with tensile straining help to prolong tensile ductility while increasing strength. These previously established strain hardening mechanisms are well exemplified in austenitic steels such as 316 series stainless steels. However, as seen in Fig. 1, even these high-performance alloys have *not* reached a combination of gigapascal yield strength simultaneously with pure-metal-like ductility approaching 50% elongation.

## 2. High-entropy alloys further improve the strength-ductility synergy

A game changer in further breaking the trade-off turns out to be the enhancement of inhomogeneity via chemical means, by going to multi-principal-element alloys (MPEAs; this term will be used in this review as an alternate to HEAs). These HEAs approach equimolar compositions, no longer distinguishing which component is the solvent and which ones are the solutes, as schematically illustrated in Fig. S3. The chemical inhomogeneity becomes ubiquitous and intensified in these concentrated mixtures. Examples include appreciable statistical fluctuations in a random solution and purposely introduced concentration undulation on nanometer scale, as well as tunable degree/extent of nanoscale LCO including but often well beyond chemical short-range order (CSRO, in the first couple of nearest-neighbor shells). These chemical inhomogeneities add a new knob to turn in microstructural design for property optimization, which is the strategy we advocate in this review article.

This strategy exploiting the chemical inhomogeneity (this wording will be used interchangeably with "heterogeneity" in this paper) takes a cue from, or naturally follows, the state-of-the-art steels, where progress has been made steadily in recent years via relatively high concentrations of alloying elements including the substitutional and interstitial (such as high-nitrogen) ones [2,14]. In these compositionally complex steels, inhomogeneities include closely spaced multiple (sometimes mutually transforming) phases [15], exploiting the massive solid solution strengthening provided by the multiple components to achieve fairly high strength and large ductility [15]. Such alloys present multi-stage strain hardening, exploiting TWIP and TRIP effects, by tuning the SFE or the metastability of the matrix phase to promote twinning or martensitic transformation. The magnitude of the SFE is important because it influences the nature of the dislocation reactions and particularly the strength of dislocation locks which promote dislocation accumulation. As will be discussed later, these strength-ductility mechanisms will be pushed further and optimized when using HEAs, with their close-to-equimolar-compositions. Moreover, the atomic size differences in the concentrated solid solution render a wider distribution of expanded and compressed interstitial sites, capable of dissolving massive amounts of interstitials [16].

There are also even more complex and ultrastrong steels with $\sigma_y$ well in excess of 1 GPa, which we will not cover in this review. This is because these alloys present additional issues, as they often rely on a significant carbon content to reach extraordinary strength. A consequence is that the carbon segregates readily and instigates localized non-uniform deformation (such as Lüders strain and PLC bands). With such a propensity towards strain localization, it is questionable to claim *bona fide* "uniform tensile elongation" at high $\sigma_y$; for example Lüders band starts at a strain of 5.9% even though the authors report uniform strain above 20% [14]. Moreover, the carbon content also leads to a considerable volume fraction of martensite, cementite or other brittle intermetallic phases that hamper ductility; bcc steels, for example, suffer from the ductile to brittle transition when the temperature is sufficiently low [17].

These drawbacks associated with complex steels, however, can be avoided by taking advantage of the recently emerging "high entropy alloys" (HEAs) [18]. For example, one can start from 316 stainless steels, which already contain a high concentration of solutes. It is just that now HEAs go one step further, with heavy alloying approaching equimolar compositions. In other words, the advent of HEAs brings an extreme version of previous concentrated solutions, which, while having compositions residing near the center of the phase diagram, are still based on relatively simple-structured fcc, bcc and hcp solid solutions. Chemical variations can then enhance inhomogeneities in a variety of "user friendly" ways to promote strain hardening, without incorporating brittle intermetallics. Note that chemical heterogeneities on the nanoscale have already been used in steels [19] and Ti alloys [20]. In fact, these steels and Ti alloys are multi-component alloys themselves, such that they may be regarded as medium-entropy alloys (MEAs), the next-of-kin to HEAs, sharing similar alloy design concepts. What is new about the MPEAs is that they further expand the opportunities when it comes to proliferating the population density of inhomogeneities with ultrafine spacing, as will be illustrated via a number of examples in this review. In this regard, local order effects are also used in many other alloy systems. For example, the Cu clustering in Al(Cu) solid solution during the early stage of GP-zone formation (forming GP zone 1 and GP zone 2 in different local composition and chemical order, etc.) is just one example. But such features are not spaced close to each other and become noticeable only after ageing for some time to form incipient GP zones. Now with concentrated HEAs, the new scenario is that we have many more (metastable) local structures, present throughout the alloy in high number population and larger volume fraction, and they are now small but closely spaced (on the scale of a few nanometers). In fact, these MPEA fcc/bcc solid solutions have inherent chemical inhomogeneities, in particular the locally varying compositions and local chemical order (including nanoscale $L1_2$ and B2 domains). It should be emphasized that the chemical inhomogeneity we discuss in this review are on the scales from sub-nanometer to a couple tens of nanometers, rather than previously familiar inhomogeneities (such as segregation during casting, in dendrite and interdendrite regions) on macro scale, or band scale on micrometer scale.

We now summarize, using the yellow-shaded area in Figure 1, the most recent discoveries over the past four years in the worldwide "gold rush" for novel HEAs (for previous cases, see our earlier review [2]). These new data points are not merely an incremental progress, but leap upwards into a territory (the yellow region in Fig. 1) seldom touched before. The latter can be appreciated from the scarlet-colored dashed line depicted in Fig. 1. Such an envelope defying the previous trend, with a shape/curvature bulging towards the best-case scenario – the upper right corner with simultaneously high strength and ductility, was targeted before using a schematic [11] (quoted in Figure S2b). We now have many HEA cases (see Fig. 1), while based on simple fcc and/or bcc crystal structures, can be stronger than upper echelon commercial alloys [3], and yet remain, surprisingly, as ductile as elemental metals. The space in Fig. 1 can now be fully utilized, without having to settle for the commonly observed trade-off. To appreciate this point, one can picture starting from the lower right corner of the box: the pure-metal-like ductility can now be retained while raising the $\sigma_y$ vertically up towards 1 GPa. Alternatively, one can start from the upper left corner: this level of $\sigma_y$ can be maintained when marching horizontally across the box. A combination of 1 GPa, 47% elongation in Fig. 1b is achievable even for bcc HEAs, for which tensile ductility has been difficult to come by [21].

Note that it is fair play to directly compare the HEAs (yellow area) with the simpler metals and alloys in the blue and pink areas, because the latter are also commonly used as the benchmark references for conventional precipitation hardening and solid solution hardening, which are traditional strengthening mechanisms with respect to the matrix elemental metal. In any event, to claim real success in strength-ductility synergy one must achieve property values that standout in previously unoccupied territory (see Fig. 1), beating specific known/standard alloys that one seeks to replace. In Fig. 1, a competitive edge over existing engineering alloys can be easily recognized for HEAs; in fact the HEAs towards the upper rim amply qualify as "strong yet ductile", which we define as "yield strength at least above gigapascal level and ductility on the par with un-strengthened elemental meals".

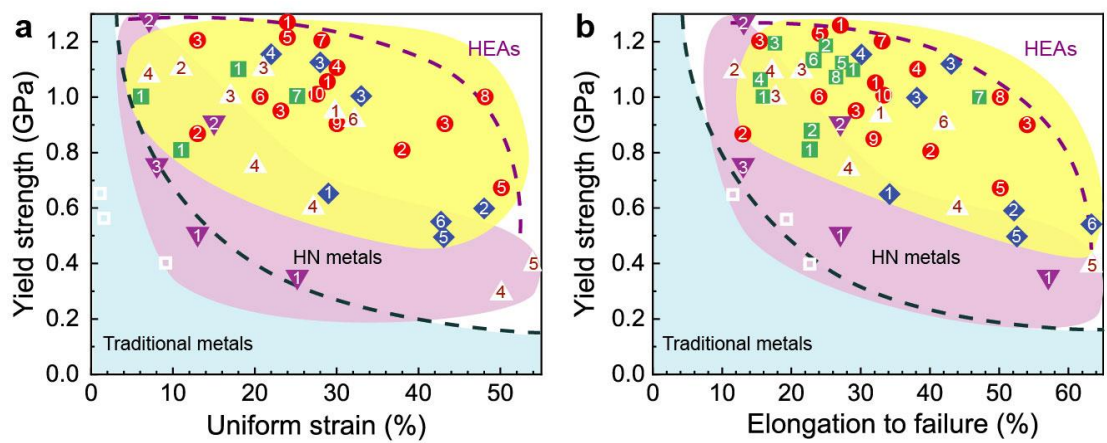

Figure 1 Yield strength versus (a) uniform tensile strain and (b) total tensile strain to failure for metallic materials. The light-blue-shaded area covers simple elemental metals and their traditional (non-concentrated) solid solutions [1], and the pink-shaded area is for heterogeneous nanostructured (HN) metals and alloys reviewed previously in Ref. [2]. Several representative Cu alloys [22] (white hollow squares) and Ti alloys (violet solid upside-down triangles) are also presented, including Ti-25Nb-0.7Ta-2Zr (at%) [20] (▽1), Ti-2.8Cr-4.5Zr-5.2Al (wt%) [23] (▽2), Ti–24Nb–4Zr–8Sn (wt%) [24] (▽3). Representative steels, including 316 austenitic stainless which are alloyed with relatively high contents of solutes and hence similar to some fcc medium-entropy alloys, are labelled using white solid triangles: Fe-0.2C–10.2Mn–2.8Al–1Si (wt%) [19] (△1), Fe-19Ni-5Ti (wt%) [25] (△2), medium Mn steel [26] (△3), 316L austenitic stainless steel [13] (△4), high-nitrogen steel Fe-17.6Cr-12.3Ni-1.19Mn-0.55Si-2.08Mo-0.012C-0.45N (wt%) [27] (△5), precipitation hardened high-

nitrogen steel Fe-18Mn-18Cr-0.6N (wt%) [28] (⬟). Most-recent HEA representatives showing superior strength-ductility synergy are found to reside in the yellow-shaded region. Several bcc single-phase HEAs are indicated using green solid squares: Ti-V-Nb-Hf [29,30] (①), Ti-Zr-Nb-Al [31] (②), N (0.6 at%)-doped TiZrNbTa [32] (③), TiZrNb(MoTa)0.1 [33] (④), O (2 at%)-doped TiZrNbHf [34] (⑤), O (2 at%)-doped Ta0.5Nb0.5HfZrTi [35] (⑥), Ti-Zr-Nb-V-Al [36] (⑦), Ti-30Zr-14Nb-3O (⑧) [37]. The fcc single-phase HEAs are represented using blue solid diamonds: N (1.8 at%)-doped FeCoCrNi [38] (◇①), (NiCoCr)$_{92}$Al$_6$Ta$_2$ [39] (◇②), VCoNi [40,41] (◇③), heterogeneous grained CoCrNi [42] (◇④), VCrMnFeCoNi [43] (◇⑤), heterogeneous gradient dislocation cell structure (GDS) Al$_{0.1}$CrCrFeNi (◇⑥) [44]. The dual-phase or multi-phase HEAs are highlighted using red solid circles: Al0.2CoNiV [45] (①), Fe$_{58.4}$Ni$_{32.6}$Al$_{6.1}$Ti$_{2.9}$ [46] (②), Fe-Ni-Co-Ta-Al [47,48] (③), Fe-26Mn-16Al-5Ni-5C [49] (④), Fe$_{20}$Co$_{20}$Ni$_{41}$Al$_{19}$ [50,51] (⑤), AlCoCrFeNi$_{2.1}$ [52] (⑥), (CoCrNi)$_{98}$Ta$_2$ [53] (⑦), (FeCoNi)$_{86}$Al$_7$Ti$_7$ [54] (⑧), Fe-Ni-Co-Ta-Al-B [55] (⑨), compositionally complex steel Fe–10Al–15Mn–0.8C–5Ni (wt%) [56] (⑩). The HEAs along the upper rim outperform previous alloys [46] in terms of strength-ductility synergy. Ultrahigh-strength steels with strength well above 1 GPa are not included for comparison in Fig. 1, because they rely on a significant carbon content to reach for extraordinary strength. The high carbon instigates not only early localized non-uniform deformation (such as Lüders band) but also a considerable volume fraction of complex phases including martensite, cementite or other brittle intermetallic phases that can severely limit ductility. The HEAs, on the other hand, are based on solutions with simple crystal structures such as fcc and bcc (and their ordered form L1$_2$ and B2). Therefore, the commercial alloys chosen for comparisons in this figure are mostly solid solutions. Austenitic stainless steels are one such example, which are multi-component fcc solutions with fairly high solute contents resembling MEAs and HEAs.

What, then, is special about HEAs in their microstructures, such that they outperform previous metallic materials and even the alloys that showed good strength-ductility synergy (see the comparison with the best prior examples in Fig. 1)? Again, for a metal or alloy strengthened to increase its yield strength (e.g., via grain size reduction or cold work that stores defects and refines the microstructure), its ductility almost always keeps decreasing with rising yield strength, from its originally large elongation before strengthening. The reason for this trade-off is the $\Theta$ failing to keep up with the $\sigma_y$ increase [1,2]. This predicament, however, is now far less prohibitive, because many HEAs are extraordinarily conducive to dynamic refinement of microstructure due to effective accumulation of more and more defects in the lattice as tensile straining progresses, such that an unusually high $\Theta$ can be sustained. Even though such a $\Theta$ may still not be as high as that for an un-strengthened coarse-grained simple fcc metal, it nonetheless is well above the low $\Theta$ normally expected for a high-yield-strength metal/alloy being plastically deformed at high flow stresses. In other words, the key point we make in this review is that alloy chemistry can be exploited to promote inhomogeneities that impart extra strain hardening. A hint in this direction can already be found in austenitic stainless steels, which are themselves concentrated fcc solid solutions much like some fcc HEAs. 316LN stainless steel [57], for example, has a higher $\Theta$, and consequently better $\sigma_y$ and ductility, than its constituent metals. Bcc Ti alloy can have high $\Theta$ and ductility through engineering nanoscale (~1 nm) ω phase, which acts as a switch between dislocation channeling and joint twinning- and transformation-induced plasticity [20]. Now that HEAs have even heavier alloying, the factors at play (and known to promote strain hardening) become more conspicuous. In Section 3, we outline the wide variety of chemically enhanced inhomogeneities in HEAs, and then explain in Section 4 the roles these inhomogeneities play in regulating dislocation behavior.

## 3. Chemically enhanced inhomogeneities in HEAs

### 3.1 Choosing alloying elements that promote embedded inhomogeneities

We begin by noting that the HEAs are inherently prone to being compositionally inhomogeneous. The statistical fluctuation (without intentionally tuning the alloy chemistry) of the composition in a given HEA is already much larger than in a normal dilute solution, even when the solution is random (and hence can be approximated as an ideal solution). This is because when the overall composition is near the center (equiatomic) region of the (multicomponent) alloy system, the standard deviation ($s$) of the concentration ($c$) is maximized (take a binary solution for example, $s = \sqrt{c \times (1-c)/N}$), where $N$ is the total number of atoms in the sample region being examined; it reaches maximum at $c$=0.5). A composition fluctuation a few at% in amplitude is therefore the norm, for local regions on the size scale of one nanometer.

This compositional fluctuation/inhomogeneity obviously would have consequences. We first look at the stacking fault energy (SFE), which is well known to be composition dependent. For HEAs, the SFE becomes a local property, highly variable from one location to another, even for a random solution [58]. This is very different from elemental metals and dilute alloys, where the SFE is a single-valued material property, unchanged everywhere in a given alloy.

As a starter in the arena of inhomogeneity design, one can judiciously adjust/select the alloying elements used in the MPEA recipe. It is well known that the strain hardening of alloys can be promoted via tuning the SFE and the metastability of matrix phases [59]. With the advent of HEAs, the SFE tends to be low, which may be perceived as due to the "correct" stacking being already a complex and compromised one in the compositionally complex solution, such that a faulted packing only incurs a small additional energy penalty. Moreover, high concentrations of alloying elements can be incorporated to push the magnitude of SFE down to single digit. Some constituent species are particularly effective in reducing the SFE. Once we know which element reduces SFE, a high content of it in an HEA, rather than as low-concentration solutes, can then render the SFE extraordinarily low (SFE not much above zero in numerous places [58]).

One example to illustrate these attributes is shown in Fig. 2, in which Wei et al. [60] substituted Mn with Si in the Cantor alloy ($Co_{20}Cr_{20}Fe_{20}Ni_{20}Mn_{20}$). This metalloid substitution reduces the average SFE from ~30 mJ/m$^2$ for the Cantor alloy to only a few mJ/m$^2$ for the Si-10 ($Co_{20}Cr_{20}Fe_{20}Ni_{20}Mn_{10}Si_{10}$) and Si-12 ($Co_{22}Cr_{22}Fe_{22}Ni_{22}Si_{12}$) HEAs. The SFE can be deduced from the splitting distance $d$ of the extended dislocation (Fig. S4). The analysis in Ref. [60] suggests that the chemical tuning lowers the magnitude of SFE to single digit. The SFE varies from location to location, as reflected by the varying $d$, i.e., the dissociation distance measured by the width of the SF bounded by the two partial dislocations that are now wavy rather than straight as in conventional fcc solutions. The very low SFE leads to a high propensity for the formation of abundant stacking faults (SFs), nano-sized hcp phases (Fig. 2a), and profuse twins (Fig. 2b) upon deformation.

Similarly low SFE is actually quite common in many fcc HEAs [61,62]. It is therefore inherently easy for deformation to induce, on the fly with tensile straining, profuse stacking faults, multiple types of deformation twins [39,50,51], (martensitic) phase transformation forming (fcc/bcc/hcp) nano-lamellae [59,62,63], (planar-slip) deformation bands that intersect [39,41,47,48], etc., all over the alloy to dynamically refine the microstructure (Fig. 2), making the alloy increasingly inhomogeneous along with tensile deformation. In an extreme case of a HEA with a near-zero yet positive SFE [62], both forward transformation (from fcc to hcp nanolaminate) and reverse transformation (from hcp phase to fcc) occur concurrently upon deformation. This enables the load-driven formation of hierarchical nanolaminate structure (Fig. 2c), which dynamically embed heterogeneities/obstacles and refine the microstructure, keeping up the strain hardening rate as will be demonstrated in Section 4.

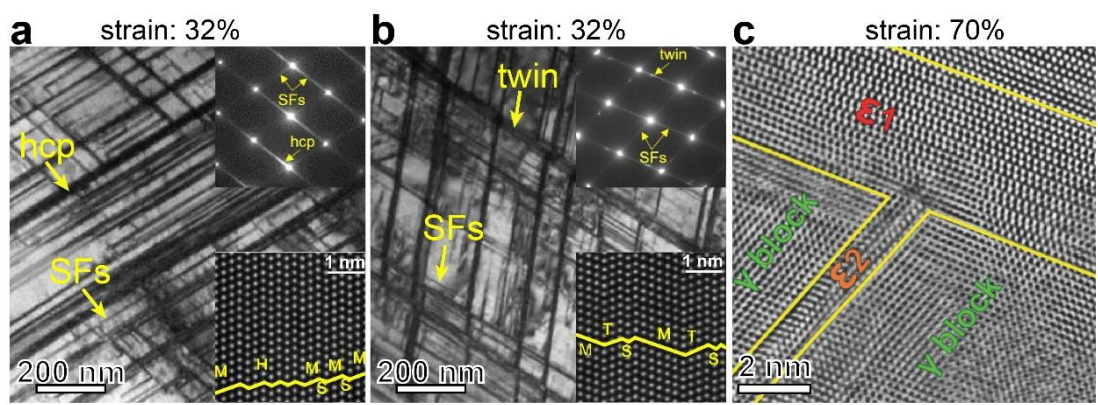

Figure 2. TEM images showing stacking faults (SFs), twins, nano-lamellae embedded all over the HEA to dynamically refine the microstructure upon deformation. (a) hcp and SFs in a deformed $Co_{22}Cr_{22}Fe_{22}Ni_{22}Si_{12}$ (at%) alloy, adapted from [60]. (b) Through changing the Si content and thus tuning the stacking fault energy, twins and SFs are formed in a $Co_{20}Cr_{20}Fe_{20}Ni_{20}Si_{10}$ (at%) alloy upon deformation [60]. The top-right insets in (a, b) are the corresponding SAED patterns, and the bottom-right insets in (a, b) are the high-angle annular dark-field (HAADF) images. The 'M', 'H', 'S', 'T' represents the fcc matrix, hcp phase, SF, and twin, respectively. (c) Bi-directional phase transformation, forward transformation from a fcc to a hcp phase or reverse transformation from a hcp to a fcc phase, enabling the formation of hierarchical nanolaminates in a $Fe_{50}Mn_{30}Co_{10}Cr_{10}$ (at%) dual-phase alloy upon deformation [62].

## 3.2 Compositional undulations in HEAs

In addition to the chemical recipe tuning in Section 3.1, The HEAs offer chemical inhomogeneities non-existent in elemental metals and conventional solid solution alloys, which are of a uniform concentration of dilute solutes. The chemical fluctuation in Section 3.1 for a concentrated solution will increase its amplitude when the solution is no longer random, and can be intentionally enlarged with wavelengths well above that in random solutions (on one-nanometer scale). For instance, Bu et al. [64] reported that their bcc TiZrNbHf alloy exhibits nanoscale local chemical fluctuation (LCF), in which Hf-enriched clusters adjoining Ti-enriched regions (Fig. 3a). Ding et al. [65] made a CoFeCrNiPd alloy, in which Pd has a markedly different electronegativity and atomic size from the other constituent elements. In the presence of Pd, compositional spikes/valleys at a periodicity of 1 to 3 nm [65] were observed (Fig. 3b).

More significant chemical fluctuations, which we call composition undulation, can be purposely installed throughout the alloy, through artificial modulation of "concentration waves" with amplitude and length scale well beyond statistical fluctuation. The electroplated Ni-Co solid solution serves as an example in this regard. Although this is a binary system and may not qualify as a HEA, it is nevertheless a close-to-equimolar solid solution and has similar features that can be expected when multiple principal elements are present. Even though the solution is expected to be homogeneously mixed (the heat of mixing of the binary system is close to zero) in a single fcc phase, it was electroplated to purposely contain obvious composition undulation on the scales of 1 to 10 nanometers (Fig. 3c) [66]. This further widens the range of the spatially varying SFE. Note that if it were a dilute Ni-Co solution, the fluctuation of Co concentration would only be a couple of at% and that only changes the SFE slightly. By contrast, in the near-equiatomic concentration regime, the SFE would almost double its magnitude when the Co concentration is designed to vary from 60 at% to 40% (see Fig. S5a and inset). Thus, in the Ni-Co alloy the SFE value undulates up and down from one place to another, spanning an unusually wide range. As a result, the moving partial dislocations would be forced to constantly adjust their splitting distance and morphology (curvature of the dislocation lines, see Fig. S4). The consequences will be further discussed in Section 4.

Such a spatial inhomogeneity in SFE is ubiquitous in many H/MEAs, although perhaps not as striking as in the Ni-Co case. Examples include fcc NiCoCr medium entropy alloys (MEAs) that are partially chemically ordered (to be discussed later), which also exhibit locally varying composition (Fig. 4a) as well as SFE (Fig. 4b). The Si-substituted Cantor-based HEA in Section 3.1 is yet another such example. It exhibits spatially varying SFE, with different segments showing different splitting distances along the (coplanar) partial dislocation line, which therefore becomes wavy, as already seen in Fig. S4 earlier.

Another way to create modulated compositional wave is through spinodal decomposition. An example is the $Fe_{15}Co_{15}Ni_{20}Mn_{20}Cu_{30}$ (at %) alloy [67] that exhibited spinodal decomposition upon heat treatment; the decomposed alloy showed improved magnetic properties with respect to the original solid solution matrix [67]. $Ti_{38}V_{15}Nb_{23}Hf_{24}$ alloy, a variant of the equiatomic TiVNbHf composition, reveals a decreased bcc phase stability and forms bct-nanoprecipitates within the bcc matrix (Fig. 3d) after fast cooling [30]. On the other hand, a different processing reported in Ref. [29] decomposed a similar TiVNbHf bcc HEA solid solution into two bcc variants with different compositions, i.e., alternating β and $β^*$ domains 50–300 nm in length and ~10 nm in width, forming a spinodal-like structural pattern (Fig. 3e). A further refined version after adding Al will be discussed later in this review. In other words, during heat treatment the thermodynamics of the alloy system drive the HEA to self-phase-separate into a compositionally undulating inhomogeneous microstructure. The magnitude and direction of the elastic distortions of the decomposed phases are now tunable, thanks to the increased degree of freedom offered by the HEA composition.

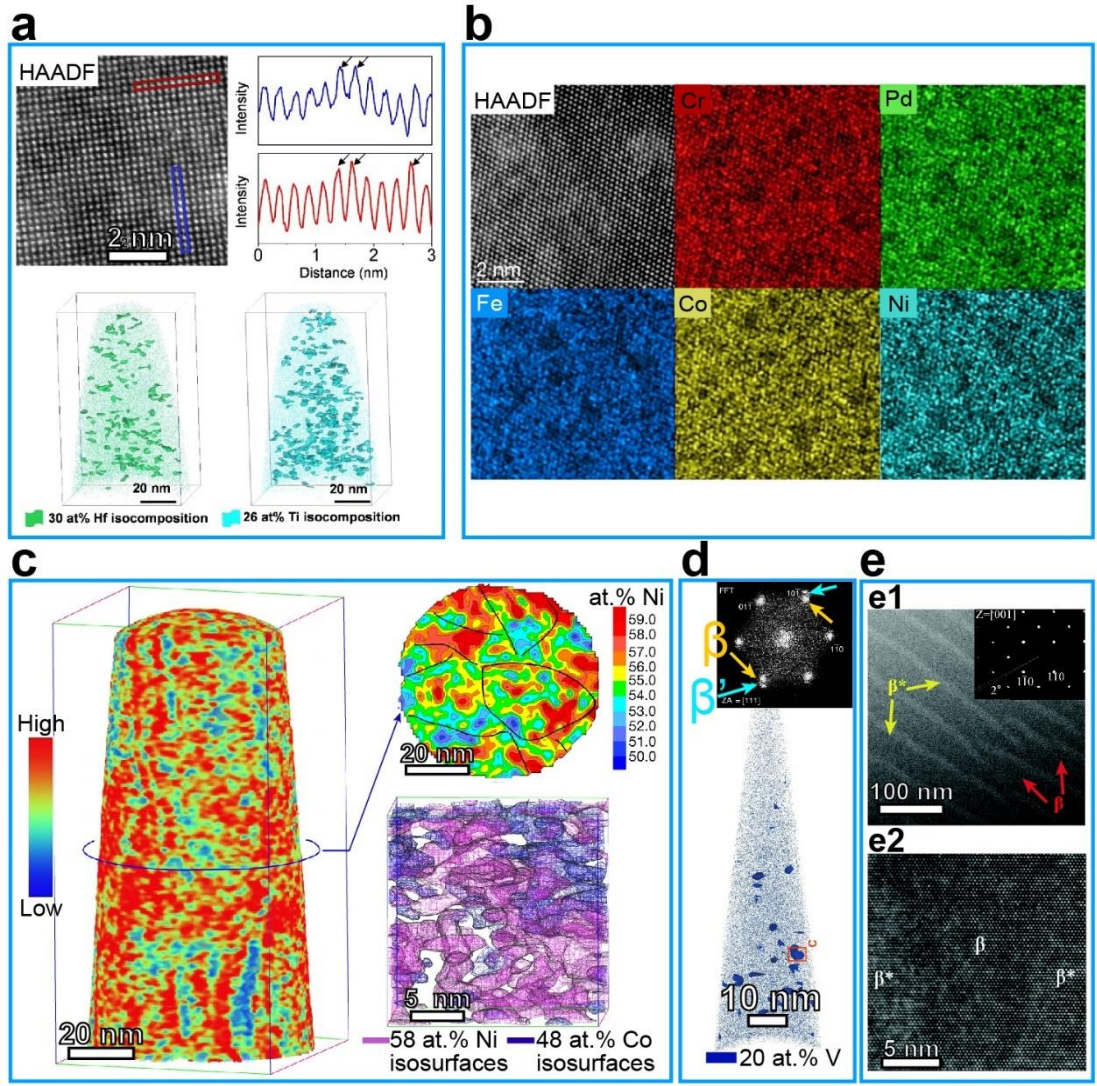

Figure 3. Single-phase HEAs with compositional undulation on different length scales. (a) Bcc TiZrHfNb alloy with nanometer scale local chemical fluctuation [64]. (b) Fcc CrFeCoNiPd alloy showing composition periodicity of 1−3 nm [65]. (c) Fcc NiCo alloy electroplated to contain composition undulations at length scales from 1 nm up to ~10 nm [66]. (d) 3D reconstruction of the APT dataset, showing V-rich nanoscale precipitates embedded in the matrix of a $Ti_{38}V_{15}Nb_{23}Hf_{24}$ (at%) alloy [30]. The top-right inset is the SAED image of the alloy, revealing overlapped patterns with a small misorientation angle, indicating bct nanoprecipitates embedded in the bcc matrix. (e) Spinodal-like modulated TiVNbHf solid solution [29]. (e1) HAADF-STEM image showing uniformly alternating β and β* phases. Inset is the corresponding SAED pattern. (e2) HAADF-STEM micrograph of a region containing β and β*. The compositional undulation wavelength can be tuned by adding Al into the HEA, which will be discussed later in this review

### 3.3 Local chemical order (LCO) in HEAs

The HEA solid solution represents a metastable middle state between the two

extremes — the random (ideal) solution state (initially of high configurational entropy) and the ground state of fully ordered intermetallics (with diminished configurational entropy). This hints at a propensity for yet another form of chemical inhomogeneity non-existent in dilute solutions, namely, variable local chemical order (LCO) [68]. Different from conventional solid solutions, in which the solute content is low and each solute atom is almost always surrounded by solvent atoms only, HEAs tend to develop LCO to various degrees/extent, from nearest-neighbor atomic shells all the way to domains of a number of nanometers in size that resemble (or, are early stages of) intermetallic precipitates. This is because of the enthalpic interaction favoring some constituent elements that inevitably come into contact in the concentrated solution, as the elemental pairs would not be identical in terms of chemical affinity. Along with the local chemical ordering, and sometimes as a result of it, chemical undulation also develops. An example of LCO developed in NiCoCr alloy is shown in Fig. 4a via atomic simulations. The constituent elements can in fact organize into various sizes of LCOs [68], depending on the alloy processing conditions (ageing temperature), eventually (if not kinetically constrained) into domains several nanometers in size. The LCOs also carry varying compositions (Section 3.1) and are inhomogeneous in the alloy when examined on the nanometer scale (Fig. 4a).

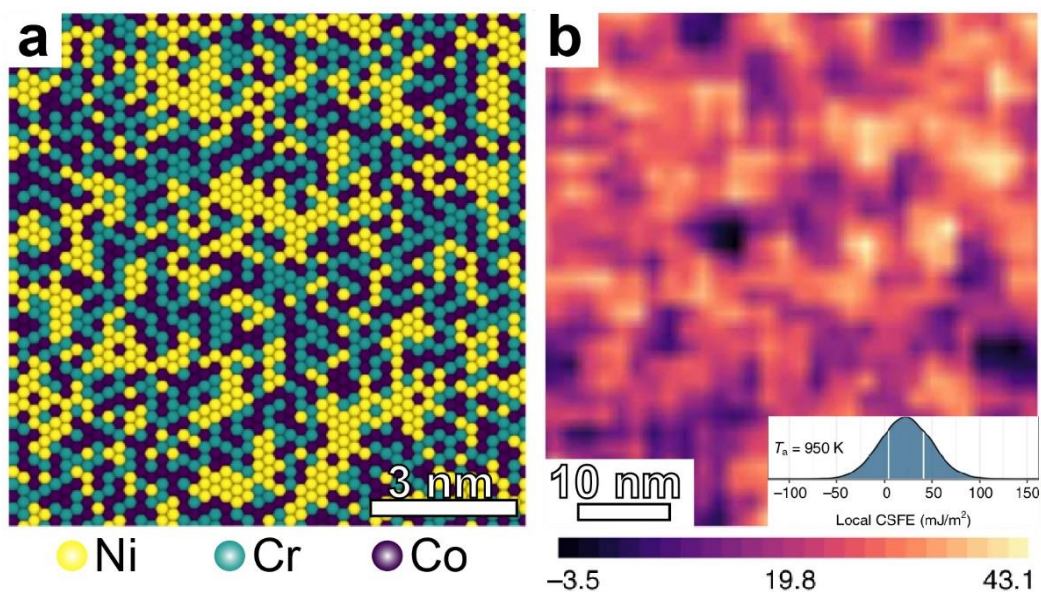

Figure 4 NiCoCr MEA aged at 950 K, adapted from the MD simulation results in [68].

(a) Representative configurations viewed on (111) plane, showing locally varying composition. (b) Spatial distribution of local SFEs. The bottom-right inset in (b) presents the (probability density) distribution of local complex SFEs. An empirical interatomic potential was used for simulating the non-magnetic NiCoCr alloy [68], which has been developed in the formalism of EAM [69,70], by matching a large ab initio database established for the ternary system without consideration of spin polarization. In order to give an accurate account of the Ni–Co–Cr system in the full compositional range, more than 3000 atomic configurations were selected to build a comprehensive ab initio database from non-spin-polarized DFT calculations. The potential was improved through an iterative process and was further refined to match experimental data, including cohesive energies, lattice parameters, elastic constants, and phonon frequencies of the constituent elements. Then hybrid MD and Monte Carlo (MC) simulations were carried out to obtain the equilibrium configurations at different annealing temperatures [68]. Sufficient MC cycles (270,000–500,000 cycles depending on annealing temperature) were carried out to achieve converged LCO to an extent that is difficult to reach in the lab due to kinetic constraints in experiments. For more details of the simulation, the readers are referred to [68].

The widespread LCOs in HEAs are to be contrasted with the case of supersaturated solid solutions in conventional precipitation-hardenable Al alloys. In the latter, the solutes are low in content and initially do not see each other. Obvious LCOs develop mainly after much ageing after precipitate precursors take shape. For example, there are stages during ageing when Al-Cu alloy would develop segregated/clustered Cu atoms and G-P zones (GP1 and GP2, e.g., single-layered to triple-layered configurations) before evolving into θ'' and θ' precipitates. Note that these LCOs are different in chemical order from one another and from the later precipitates (the θ'' and θ'), each having its own spatial dimensions, atomic packing configuration, local composition and degree of chemical order. Such LCOs are therefore difficult to identify and classify, although they could be vaguely regarded as metastable early-stage precursors of precipitates. Now that the HEA is a highly concentrated solution to begin with, we have more types and sizes of (metastable) local structures, present throughout the alloy in high number population and large volume fraction, starting out small but being closely spaced (on the scale of a few nanometers).

The earliest stage of LCOs in HEAs begins with chemical short-range order (CSRO), which refers to the nearest-neighbor preference or avoidance of some of the constituent elements. Such CSROs arise from subtle and partial ordering and are arguably the most difficult to decipher. Chen et al. [71] recently analyzed the CSRO in fcc VCoNi medium-entropy alloy (MEA), and revealed that the CSRO is composed of two V-enriched ($\bar{3}11$) planes sandwiching one CoNi-enriched ($\bar{3}11$) plane. This chemical information complements diffuse electron reflections that may correspond to the 1 nm-scale CSRO regions in real space. Chemical medium-range order (CMRO) is the next level beyond the CSRO motif, being larger in size (~2 nm) than that of CSRO (<1 nm in at least one of three dimensions) and as such may begin to possess a specific (such as CuPt-type $L1_1$) symmetry [72]. Wang et al. [72] demonstrated that both CSRO and CMRO emerge in a dual-phase (fcc-B2) $Al_{9.5}CrCoNi$ alloy (Fig. 5a-c), showing two Cr-enriched ($\bar{3}11$) planes sandwiching one Cr-depleted plane, different from the chemical order of the known (B2) second phase at this alloy composition. "Short-range-order" domains up to ~ 5 nm in size have also been reported for a CrCoNi MEA [73]. While such reports call all small domains "short-range order", we put it in quotation marks here because "short-range" could cause confusion, as 5 nm is typical for nanocrystals, and all crystals are long-range ordered by definition, conflicting with short-range order. For periodic chemical patterns on such length scales, LCOs would be a more appropriate term. LCO starts from CSRO but can also cover partially to fully ordered metastable domains up to a number of nanometers in (at least one of the three) dimensions.

A most recent example of CSRO/LCO is in the bcc $Ti_{50}Zr_{18}Nb_{15}V_{12}Al_5$ multicomponent alloy. At this composition its equilibrium precipitate is a $Zr_5Al_3$-type intermetallic phase [36], but upon annealing this MPEA the initial chemical order developed is a B2-like LCO, as shown in Fig. 5d-f, and in Fig. S6 from a comparison between the projection of a B2 model with the HAADF images showing alternating Zr-enriched and Al-enriched atomic planes [36]. Note that in this bcc case (bcc structures are known not to form SFs due to high energy barriers) the extra diffuse disks in the electron diffraction pattern cannot arise from tiny planar defects suspected to co-exist with CSRO in fcc HEAs [71,73]. In all cases, chemical information showing the local atomic packing preference/avoidance (Fig. 5) is needed as a tell-tale indicator to lend support to the presence of LCO.

In addition to LCO constituted by principal elements, one can add interstitial solutes into the HEA to help promote chemical ordering. Jiao et al. [37] found that doping oxygen into a bcc Ti-Zr-Nb alloy increases the degree of LCO, rendering the formation of CSRO, CMRO, all the way to even longer-ranged chemical order. In fact, adding interstitial oxygen (or other chemically distinct species such as N and C) that bonds strongly to certain species such as Zr and Ti to form complexes [34] is an efficient route to promote LCO, which by the way also comes with associated local compositional undulation. In the Ti–30Zr–14Nb case, the interstitial oxygen solutes are accommodated in the form of ordered oxygen complexes (OOCs enriched in O and Zr/Ti): a high number density ($8.7 \times 10^{24}$ m$^{-3}$) of OOCs was formed with the addition of 3 at% of oxygen (Fig. S7). The clustering of solutes would be akin to previously known Cottrell atmosphere and Suzuki segregation, but now with solute-promoted local compositional undulation and chemical ordering involving the principal elements in the HEA.

The expanding chemical ordering still remains LCOs if their sizes are limited to local regions measuring no more than a few nanometers in at least one out of the three dimensions. Nonetheless the LCOs could ultimately evolve into fully ordered equilibrium intermetallic precipitates, such as a L1$_2$-type (chemically ordered fcc) domains [54], each of which is expected to impact properties more significantly than the earlier LCOs including the tiny CSROs, as will be discussed separately in the next sub-section (Section 3.3).

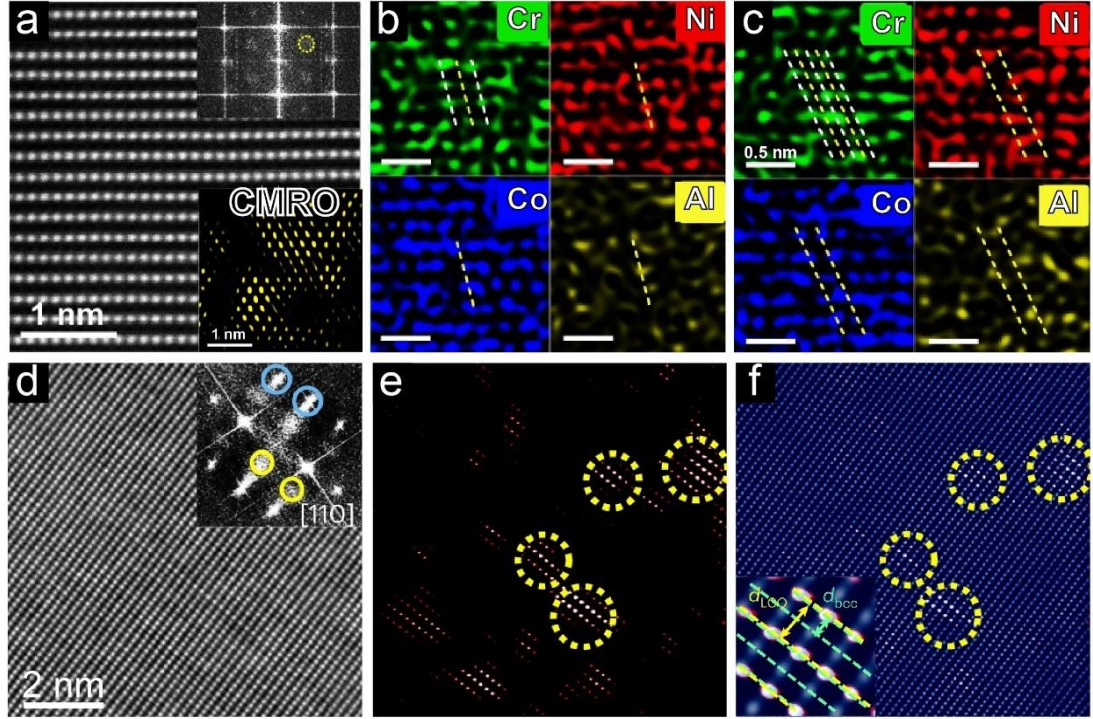

Figure 5. Local chemical order in fcc and bcc HEAs. (a-c), Chemical short-range order (CSRO) and chemical medium-range order (CMRO) in a $Al_{9.5}CrCoNi$ (at%) alloy [72]. (a) HAADF image under [112] zone axis. The top-right inset is the corresponding fast Fourier transformation (FFT) pattern, the bottom-right inset is the inverse FFT (IFFT) image showing CMRO regions. (b and c) EDS maps across an individual CSRO region in (b) and CMRO region in (c). The dashed yellow/white lines highlight Cr-depleted and Cr-enriched $\{\bar{3}11\}$ planes. (d-f), LCO in a bcc $Ti_{50}Zr_{18}Nb_{15}V_{12}Al_5$ alloy [36]. (d) HAADF image of the alloy under [011] zone axis. The top-right inset is the corresponding FFT pattern. The blue solid circles indicate the diffraction spots of the bcc matrix, and the yellow solid circles highlight the extra discs due to the B2-like LCOs. (e) IFFT image obtained from the extra disks of the FFT pattern, with the yellow dashed circles highlighting the LCOs. (f) Image superimposing the LCO (in e) and bcc IFFT images. The inset in (f) is a close-up view of an LCO region, where $d_{bcc}$ is the spacing of the {001} planes in the bcc lattice, and the $d_{LCO}$ is the spacing of the extra LCO.

Before ending this sub-section, we note that it is not imperative for local ordering and concentration undulation to take place at the same time. Chemical ordering may happen and distribute across the alloy in a non-uniform fashion without causing composition undulation: imagine the formation of local γ' domains with $L1_2$ order inside a γ fcc solution, all with $Ni_3Al$ composition. Conversely, compositional undulation can also happen alone without being accompanied by chemically driven ordering, as shown in the example in Section 3.2, which is a "random" Ni-Co solid solution with nearly zero heat of mixing. However, in such solutions each atom in fact still differs in its environment (chemical make-up of the first neighbors) because of the appreciable composition fluctuation in the concentrated solution, such that the calculated short-range order parameter $\alpha_i$ for many atoms would be different from that of a truly random solution of the sample-wide composition (although the heat of mixing is zero so there is no chemical driving force for ordering). In this sense/context, a real-world highly concentrated solution would not be "completely chemically disordered everywhere", i.e., free of CSRO.

### 3.3. Abundant interphase interfaces in MPEAs

The upper bound of chemical inhomogeneity is when a HEA phase-separates into simple-structured lamella/domains of chemically ordered fcc/bcc. Unlike in a conventional (host-solute) alloy, where precipitates are relatively few and located far apart in the matrix, when dual/multiple phases co-exist in an HEA each of them is of an unusually high volume fraction, with precipitate sizes and spacings on the scale as small as a few nanometers. For the sake of convenience, here we still refer to the coexisting phase as "precipitates" as if a second phase has nucleated in the host, although it is actually all over the place and may occupy as much space as the matrix phase. Examples include (often coherent) intermetallics particles [56] having $L1_2$- (chemically ordered fcc) and B2- (chemically ordered bcc) structures (Fig. 6a,b) [46,48,50,51], fcc/bcc/hcp [59] variants that undergo martensitic transformation from one to the other during deformation, as well as dual-phase eutectic (Fig. 6c,d) and other lamellae structures [50,51]. Furthermore, dual-precipitation of nano-sized shearable $L1_2$-type $\kappa$-carbide particles and non-shearable B2 particles (Fig. 6e) was found in a Fe-26Mn-16Al-5Ni-5C (at%) alloy [15]. Dual-heterogeneous structures have been produced in a $Co_{34}Cr_{32}Ni_{28}Al_3Ti_3$ (at%) alloy [74], which is composed of a heterogeneous fcc matrix with both coarse grains (10−30 μm) and ultra-fine grains (0.5−2 μm), together with heterogeneous ($L1_2$-structured) nanoprecipitates that range in size from several to hundreds of nanometers (Fig. 6f). These dual-precipitation or dual-heterogeneous structures constitute a very high population of interphase interfaces to cause moving dislocations to stall or pile up [50,51]. Their roles in promoting strain hardening in addition to strengthening will be discussed in Section 4.

Note that even though high concentrations of alloying elements are employed in the HEAs, brittle intermetallics usually do not reign, because the HEAs are by and large based on fcc or bcc solid solutions and their chemically ordered ($L1_2$ or B2) counterparts. Phase transformations rarely involve complex crystals with a low-symmetry unit cell containing a large number of atoms, and the interphase interfaces are often coherent and resistant to decohesion. Take a fully $L1_2$-structured $Ni_{43.9}Co_{22.4}Fe_{8.8}Al_{10.7}Ti_{11.7}B_{2.5}$ (at%) superlattice alloy [75] as an example, its grain boundaries are replaced by fcc solid solution nanolayers that are coherent with the host grains. This fcc solid solution nanolayer enhances dislocation mobilities and prevents intergranular fracture; the resultant "superlattice alloy" thus attains a high yield strength (1.0 GPa) and good tensile ductility (25%) [75] (although not as impressive as the 50% when an ultrahigh density of nanosized ductile intermetallics $L1_2$ was embedded in the fcc matrix [54]). In addition, segregation of solutes such as oxygen at the grain boundaries, which harms ductility via grain boundary fracture [37], is controlled at a low level as well (e.g., in the Ti–30Zr–14Nb-3O alloy [37]). For instance, the NbMoTaW HEA is sensitive to oxygen contaminants that weaken grain-boundary cohesion to add initiation sites for flaws and cracks. This problem can be alleviated by addition of boron, which preferentially replace oxygen at grain boundaries, changing the intergranular fracture to transgranular fracture [76].

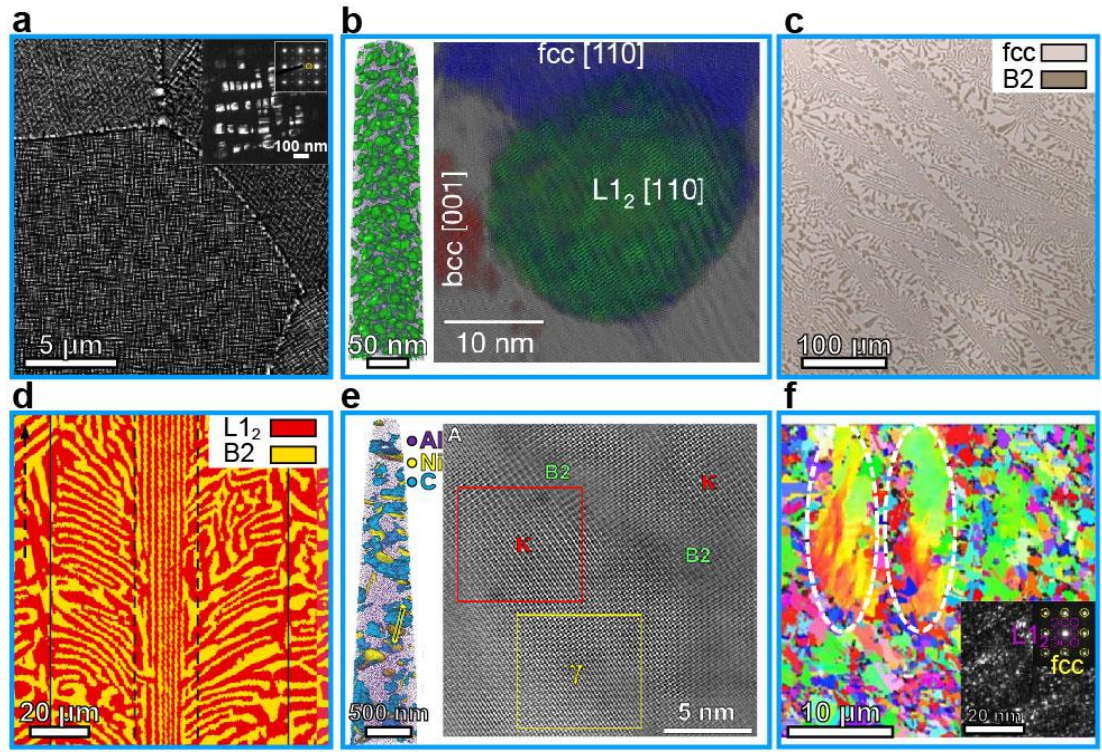

Figure 6. HEAs with high volume densities of co-existing phases ("precipitates" in a matrix). (a) Fe–32.6Ni–6.1Al–2.9Ti (at%) alloy showing nanoscale precipitates (L1$_2$ structured) and bcc phase embedded in the fcc matrix [46]. (b) Nano-sized L1$_2$ precipitates, which are coherent with the matrix, are uniformly distributed in a fcc solid solution [47]. (c) AlCoCrFeNi2.1 (at%) eutectic HEA, with B2 embedded in a fcc matrix [77]. (d) Al$_{19}$Fe$_{20}$Co$_{20}$Ni$_{41}$ (at%) eutectic HEA with a hierarchically organized herringbone structure [50]. (e) Fe-26Mn-16Al-5Ni-5C (at%) alloy reveals dual-precipitation of nanosized L'1$_2$ κ-carbides and NiAl-type B2 particles [15]. (f) Co$_{34}$Cr$_{32}$Ni$_{28}$Al$_3$Ti$_3$ (at%) alloy showing a dual-heterogeneous structure, which is composed of a heterogeneous fcc matrix with both coarse grains (from 10 to 30 μm) and ultra-fine grains (0.5 to 2 μm), together with heterogeneous nanoprecipitates (L1$_2$-structured) ranging from several to hundreds of nanometers [74].

To recapitulate, in this Section 3 we have seen that the highly concentrated HEAs usher in spatially varying compositions and LCOs, which can be tuned in degree and extent, all the way from nearest-neighbor atomic shells to "second-phase" domains. These various chemical inhomogeneities are on nanometer length scales, well below those of the familiar cast inhomogeneities. We mention in passing here that one can further push the envelope of both chemical and structural inhomogeneities via additive manufacturing, by exploiting tunable 3D-printing parameters. For example, in an additive manufactured AlCoCrFeNi$_{2.1}$ eutectic HEA [78], L1$_2$ and B2 phases coexist to install abundant interfaces, and each phase shows pronounced compositional inhomogeneities inside (Fig. 7a, Cr-rich nanoprecipitates 3−10 nm in size are formed in the B2 phase). Such 3D-printing [52] can also tailor-make a hierarchical structure (Fig. 7b) in the form of dual-phase nanolamellae embedded in microscale eutectic colonies. Besides, the B2 nanolamellae contains compositional modulation on a nanometer scale similar to spinodal decomposition. Such eutectic HEAs [77], and in particular the 3D-printed versions of them [52,78], therefore offer hierarchical inhomogeneities over a range of length scales, which, with mechanism to be discussed in the next section, promote the synergy between yield strength and tensile ductility (typical representatives [52] have been included in Fig. 1).

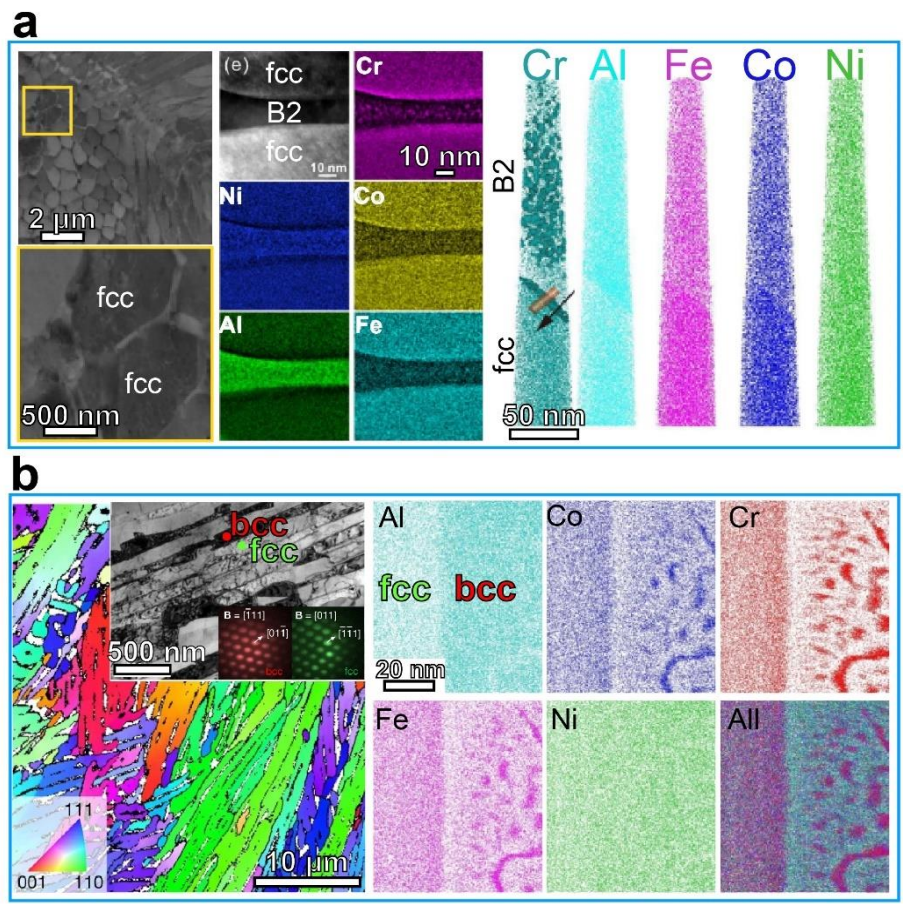

Figure 7. HEAs containing abundant heterogeneities in terms of both chemical and structural features. (a) Fcc-B2 AlCoCrFeNi$_{2.1}$ eutectic HEA prepared via additive manufacturing [78], in which Cr-rich nanoprecipitates are homogeneously distributed in the B2 phase. (b) Nanoscale composition modulation in an additive manufactured AlCoCrFeNi$_{2.1}$ eutectic HEA [52].

## 4. Chemical inhomogeneities facilitate strength-ductility synergy

Having demonstrated in Section 3 the various levels and examples of inhomogeneities inside the HEAs, we now use a separate section to focus on an explanation of the mechanisms responsible for the strength-ductility synergy promoted by the HEAs. Highly concentrated solid solutions, unlike dilute ones with almost-fixed composition and local chemical environment throughout the alloy, offer tunable degree of chemical heterogeneities, including locally varying compositions that are patch/stripe-like, all the way to fully ordered domains (L1$_2$ for fcc and B2 for bcc) that have dimensions and spacings on the scale of nanometers. Such a mixture, while retaining a rather simple crystal symmetry, can be likened to a "nano-cocktail" solution, which we define as a mixture of nanometer-scale domains with spatially varying composition and often highly diffused interfaces.

### 4.1 Strength elevation

This sub-section first illustrates the origin of the high strength in HEAs. This topic has been dealt with in many papers and reviews [79–81]; we therefore limit ourselves to a general picture while stressing the role of the inhomogeneities.

To begin with, it is easy to understand that the compositionally varying nano-sized patches/stripes lead to environment-specific dislocation segments that are a few nanometers in characteristic length scale. In front of a dislocation line, the local inhomogeneities can be perceived as producing a roughened landscape for its propagation through the lattice. In general, a HEA can be regarded as an extremely-refined cocktail solution composed of multiple ingredients that spatially mix on atomic to nanometer length scale for a random solution, and form stripes/patches several nanometers in dimension when a hierarchy of compositional wave/order is introduced. In fact, even when a highly concentrated solution is deemed to be random, it is inherently inhomogeneous, because there will be composition fluctuations much larger than in dilute solutions, as mentioned earlier in Section 3.1: the standard deviation characterizing the statistical fluctuation of concentration is maximized in the equiatomic composition range. When undulating compositions (with changing lattice constants) and different LCOs are made present, the HEA can be a mixture with continuously varying chemistry and local order/symmetry. In the extreme case the inhomogeneities can become well-defined domains that are spaced unusually close. In any case, the dislocation navigates in a choppy sea crossing inhomogeneous composition, chemical order, SFE, and diffuse interphase boundaries (or even discrete interface such as fcc/bcc). In other words, the solution is full of roadblocks, i.e., "speed bumps", which would constrain and frustrate dislocations. This naturally increases the flow stress [68]. The dislocation marches via forward slip of nanoscale segments, each escaping from its locally favorable environment. This results in "stop-and-go" (Fig. 8 and Fig. S5b), in a new variant of stick-slip different from that in conventional solid solutions (such as at L-C locks), with a corresponding activation volume/area on the length scale of several nanometers, far smaller than that of hundreds of Burgers vectors in normal fcc metals. This nanoscale-cocktail strengthening mechanism has been coined as "nanoscale segment detrapping" (NSD) [68,82]. The moving dislocation line becomes wavy (Fig. S4) because each segment responds differently as it is not in the same local environment as the adjoining ones. This picture is depicted in Fig. 8 and Fig. 9, where each NSD process involves a segment of dislocation line residing in a local region different from others, both the segment and local region being a few nanometers in terms of their characteristic length scale. In Fig. 8, the local environment in front of the moving dislocation line varies mainly due to statistical fluctuation of composition (the model is a random solution only having a low degree of LCO, i.e., negative $\alpha^1$ parameter for Co-Cr short-range order, due to compositional fluctuation on the atomic scale; if the degree of LCO is increased, the ordered patches would become stronger roadblocks to the dislocations cutting through. The activated nanoscale segments are seen to detrap from some of the local regions (as denoted by arrows). Some segments have already moved while others lag behind, producing a wavy dislocation line (continued bow-out increasing the curvature is counteracted by the self-energy of the dislocation, i.e., line tension). In other words, as long as there is chemical inhomogeneity (even just fluctuation in composition, which may be accompanied by a bit CSRO), obvious waviness of the dislocation line and its NSD forward motion will result, a detailed view of which is given by the atomistic simulation shown in Fig. 9. The resultant retardation on dislocation motion in H/MEA will be more significant when compared to elemental metals or dilute solutions. This new NSD hardening mechanism with segments exhibiting stick-slip is markedly different from traditional solid solution hardening, where the dislocation line remains straight and moves as a whole through the dilute solution, because it only needs to overcome the dragging from the relatively weak strain field of individual atoms that are few and far between, one single solute atom at a time, as shown in Fig. S5b. Also, the motion of dislocations in inhomogeneous concentrated solutions is far more intermittent than in homogeneous solid solutions. We see clear "stop-and-go", which is what we mean by stick-slip (yellow curve, Fig. S5b). This is different from the dislocation movement in a truly random solution without LCO and chemical fluctuation, or a simple fcc metal. In the latter cases, the drag on the moving dislocation is much weaker and the "hold" is of a much shorter duration, such that the distance vs time plot remains linear, giving rise to an almost constant dislocation speed (blue and orange curves in Fig. S5b).

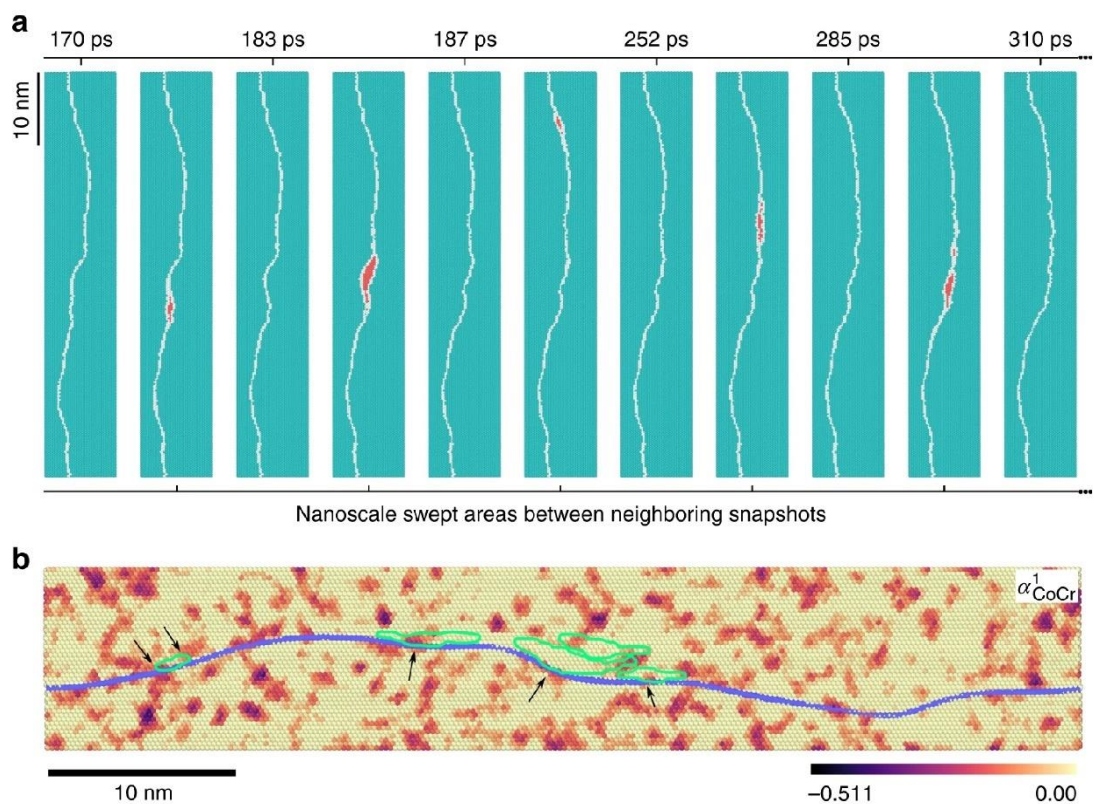

Figure 8. Nanoscale segment detrapping (NSD) in a NiCoCr alloy [68]. (a) Forward motion of dislocation line, simulated at a shear stress of 50 MPa and 300 K. The swept areas in different NSD events are highlighted using red color. (b) Correlation between NSD events (the swept areas in (a) are now outlined in green) and their inhomogeneous local environment. The chemical inhomogeneity is mainly composition variations due to statistical concentration fluctuation, because the solution is a "random" model (without enthalpy-driven chemical ordering upon ageing) on average. The composition fluctuation necessarily instigates a low and varying degree of CSRO, as the make-up of the first-neighbor shell changes from atom to atom in this concentrated solid solution (each atom is colored according to the chemical short-range order parameter $\alpha^1$ for Co–Cr). The inhomogeneous composition (with or without LCO) necessarily introduces a spatially varying SFE and shear-fault energy. Nanoscale dislocation segments detrap from energy favorable regions. As such, the NSD stick-slip mechanism (also see Fig. S5b) arising from the chemical inhomogeneity (which can be intensified when interacting with heightened LCO [58], a zoom-in view of which is shown below in Fig. 9) is different from traditional solute hardening in dilute solutions (see text).

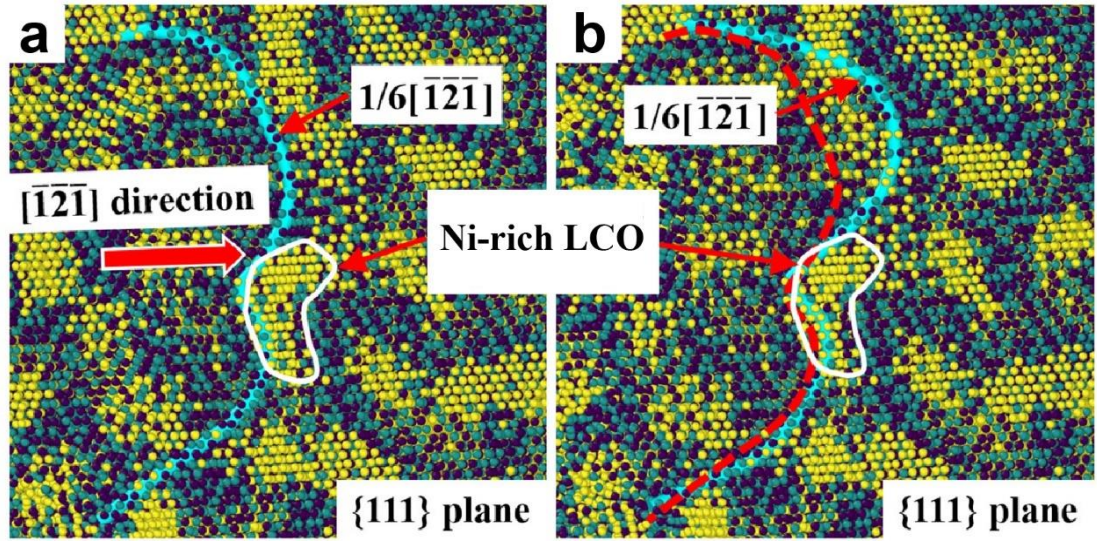

Fig. 9. Typical dislocation propagation mechanism in NiCoCr MEA, adapted from [83]. This enlarged view highlights dislocation lines (in color) moving from left to right in a wavy and jumpy mode, navigating through a chemically varying lattice (such as that in Fig. 4). As seen in (b), the original blue line in (a), now red line in (b), does not move as a whole, as some segments are halted temporarily when confronting certain local regions enriched in Ni (an example of this LCO is circled in white). In this example, the segment above the LCO has moved forward to the new blue position. The width of the box in (a) is approximately 12.5 nm.

The new strengthening mechanism can be pictured as due to the presence of a persistent dragging force imposed on dislocations moving through the "nano-cocktail" solution. Even for a solution close to being random, the dislocation traversing the distance ($L$) runs across an uneven field, constantly fluctuating in local composition and chemical order. In terms of energy ($E$), the landscape looks like waves with valleys and hills, even when not necessarily stepped planes as in the more heterogeneous cases with distinct domains /patches. This includes the fcc case, where an extended dislocation always has to dissociate into partials, splitting into variable width (a SF bounded by partials with a distance commensurate with the local SFE). In any event, the moving dislocation in concentrated alloys always experiences a finite gradient, $dE/dL$, which implies an extra dragging force that counteracts the externally imposed stresses. Putting it another way, the dislocation being driven through invariably feels a "hand" trying to "hold" it in its energy-favorable locations. As such, additional stresses are needed to dislodge the trapped dislocation segment to continue its forward motion. The NSD entails a sizable barrier corresponding to the energy expense (on the order of a fraction of one eV [58]) such that the "nano-cocktail" hardening (Fig. 7 [68]) due to collective trapping is far more significant than traditional solid solution hardening, where the trapping is provided by the strain field of just a single solute. In this context, the "stick-slip" in concentrated solution is analogous to atmosphere pinning as in strain ageing. Note the drag force (*dE/dL*) on the moving dislocation segment, which has to dislodge from energy favorable positions again and again, is almost ubiquitously present across the whole alloy. This is different from dilute solution, where the dragging happens only at isolated solute atoms, which are few and far between.

Modeling to predict the collective strengthening in concentrated solutions dates back to Labusch [84] with the underlying characteristic scales recently quantified by Leyson and Curtin [85]. The key quantity in the theory is the change in energy of a straight dislocation segment of length $\zeta$ as it glides a distance $w$ through the random solute field. The total dislocation energy is determined by an interplay between the potential energy, approximated by elastic strain energy using a concentration-weighted mean-square misfit volume, and the line tension (elastic energy due to the change in line length during the advancement) of the segment taking a wavy configuration. This total energy is minimized with respect to possible scales of fluctuation ($\zeta$, $w$) for a given line tension to obtain the controlling lengths $\zeta_c$, $w_c$, which are the characteristic length scales for the collective concentration fluctuations to create the dominant energy barrier controlling the yield stress. For NSD, the trapped segment $\zeta_c$ needs to be pushed out of a local potential energy well, and the predicted shear stress aligns well with measured or molecular-dynamics simulated ones. The misfit volume strengthening was later modified by Antillon et al. to incorporate the CSRO effects [80], accounting for the energy change when specific chemical bonds are broken. An (diffuse) anti-phase boundary strengthening term can also be added, if the HEA starts out as a solution with an appreciable degree of chemical order.

Different from such solid-solution strengthening models, in which the parameters are explicitly the elastic modulus and size misfits of alloying elements, Ngan's recent model provides a different perspective [79], which echoes our viewpoint that the spatial inhomogeneities of the alloying elements and chemical ordering can be perceived as the root cause making the dislocation motion difficult. With the inherent inhomogeneities, the energies of the planar faults created by the dislocation shear are locally varying [58] over the slip plane. The model considers the work done during dislocation advancement, against the shear-fault energy field $\gamma$ (x,y). A dislocation in local regions enriched in hard atomic motifs would be pinned, as its further advancement would produce faulting with elevated energy. This fault-energy fluctuation resistance then controls the detrapping and therefore the strength, and is a cover-all parameter as the change in fault energy reflects all the elastic and chemical interactions that affect the dislocation, including its shape and line tension. Their simulations of the local fluctuations of the energy of CSRO shear-faulting in the NiCoCr example showed that the magnitude of this form of dislocation resistance is in good agreement with strengths predicted from molecular dynamics simulations and experiments.

From the picture above, the strengthening provided by a concentrated fcc solution is dominated by the random alloy strength, as the CSRO strengthening would not last long enough to dominate macroscopic slip because successive dislocation glide would destroy the subtle ordering. For chemical inhomogeneities to elevate strength well beyond random solution expectations, it seems not enough to have just CSROs (i.e., chemical ordering in the first couple of atomic shells), or only compositional fluctuation with a spatial correlation length approaching the statistical fluctuation in a random case). Also note that if the HEA is a substitutional solid solution, the development of CSRO usually means two dissimilar species tending to be nearest neighbors. This pairing largely cancels out the lattice distortions that each of the two would brought into the lattice. The hardening effect introduced by CSRO is then offset by the simultaneously diminished elastic strains [36], or in other words by the reduced elastic misfit as a key strength contributor in Curtin's model [86]. Another way to put it, is that in this context adding CSRO would not get to increase the waviness of the dislocation line and the shear-fault energy fluctuation to increase the dislocation line tension and hence strength [87]. Taken together, all these explain why Yin and Curtin [88] observed that the yield strength of CoCrNi can be largely accounted for by their random solution model, in the absence of an analysis of the CSRO effects. In lieu of CSRO, our viewpoint emphasized and advocated LCO for strengthening [58], i.e., robust chemical correlation with length scales up to a few nanometers, such that the anti-phase boundary energy [58,80,87] associated with chemical ordering becomes an important difference that needs to be added as a separate term contributing to the strength [80]. Strengthening from LCOs would not subdue easily as CSROs because, while they are still local such that dislocations cut through them, they are nevertheless nanometers in dimensions and better ordered to survive repeated dislocation slip, and as such can act as a main contributor to strengthening akin to precipitates. The robustness of the LCOs over CSROs is another reason why we do not categorically equate LCO with CSRO. In other words, not only the extent and degree of chemical order in the former are not the same as in the latter (see Section 3), their potential consequences are also different when it comes to affecting dislocation actions. LCOs and concentration undulations on the scale of a few nanometers, going beyond atomic-scale CSROs and matching the $\zeta_c$ characteristic of NSD segment in the nanoscale cocktail solution, are expected to be more effective and persistent in offering additional resistance above Peierls stress (or lattice friction in metals or often-dilute solid solutions presenting rather flimsy obstacles) to dislocation propagation, driving up the curvature and the tension of the stick-slipping dislocation line, beyond what is already the norm in a concentrated HEA (as discussed with models in the preceding paragraph).

## 4.2     Promoting strain hardening to sustain tensile ductility

A main focus of this review is to make it transparent as to what the HEAs have to offer such that the tensile flow is better stabilized at high stresses, enabling the extension of elongation. It is perhaps less obvious as to what impact the chemical inhomogeneities would have, in terms of raising strain hardening rate and hence tensile ductility. We therefore devote the discussion in the rest of Section 4 to clarify this key issue. Our insight is that the chemically heightened ruggedness of the atomic/energy landscape is highly influential on the evolution (tangling, reaction, multiplication, accumulation) of dislocations, profoundly affecting the shape of the stress-strain curve (strain hardening), even more so than the initial plastic flow near yielding (e.g., the yield strength). To systematically explain the effects of chemically enhanced inhomogeneities on strain hardening, we start with a revisit to the case in Section 3.1. Recall that substituting Mn using Si in the Cantor alloy reduces its SFE to only a few $mJ/m^2$. The low SFE promotes the formation of abundant SFs, nano-sized hcp phases (Fig. 2a), and twins (Fig. 2b) upon deformation. The population of these embedded defects continuously rises as the flow stress increases, dynamically refining the microstructure on the fly with tensile deformation as the partial dislocation mediated processes (such as the formation of stacking faults and deformation twins) keep adding debris. The large separation distance of the partial dislocations due to the low SFE also retards their dynamic recovery. Hence, the dislocation density continuously rises to exceedingly high levels at high strains (beyond 60%, Fig. S8) [60], much more so than the base Cantor alloy. The dynamic refinement of microstructure sets roadblocks for dislocations and enhances dislocation interactions, entanglement, multiplication, and storage, leading to efficient defect accumulation. Compared with the base Cantor alloy (CoCrNiFeMn), which behaves similarly to 316 stainless steels, the chemical tuning with high content of Si-10 and Si-12 effectively enhances the inhomogeneities in the HEA, which therefore exhibits an elevated strain hardening rate (Fig. S8a) and hence synergistically rising strength and ductility (Fig. S8b), with no trade-off to speak of. The data point in fact goes off the chart (beyond the limit along the x-axis in Fig. 1).

We next discuss the notable role of LCO and compositional undulation in promoting strain hardening. Note that even the very small CSROs have associated local elastic strains [71]. The dislocation also has tensile and compressive regions at the dislocation core. Hence, dislocations tend to stall at locations where CSROs are present, such that when strains of opposite sign overlap, part of the strains cancel out to lower elastic energy, as shown in Fig. S9. It would cost additional stresses to detach the dislocation from the CSROs. This is expected to contribute to the accumulation of dislocations and hence strain hardening.

In a nutshell, the various types and degrees of chemical inhomogeneities discussed in this article render the dislocation motion in HEAs sluggish and jumpy, yielding the following consequences on strain hardening. First, a distinctive feature of dislocations in HEAs is that they are highly wavy when at rest or traversing along the tortuous trajectory, as shown in previous simulations [89,90] and experiments [91,92]. Second, the "stop and go" stick-slip offers extra wait/dwell time (Fig. S5b) such that moving dislocations are more likely to hit stalled ones. Third, the wavy morphology increases the total line length to also increase the chances for dislocations to tangle with each other. All these increase the chances for dislocations to run into each other and react, promoting dislocation multiplication and accumulation that imposes long-range force resisting ensuing dislocations. Strain hardening rate becomes elevated as a result, which sustains tensile ductility.

### 4.3 Prominent examples of chemical inhomogeneity effects

Chemical inhomogeneities can be promoted by incorporating suitable alloying element, specifically those that are chemically quite different from the principal elements of the HEA solution. The first type of added alloying elements are non-metallic solutes that take interstitial sites. An example is the work by Lei et al. [34] serves as an example that quantitatively demonstrates the effectiveness on strengthening as well as strain hardening. The oxygen interstitials they incorporate into a bcc Ti-Zr-Nb alloy preferentially hold on to Zr and Ti to form nanometer-sized LCO, which also causes associated local compositional undulation. These chemical inhomogeneities were called "ordered oxygen complexes" by the authors [34]. These OOCs do not readily dissociate (or they may easily reform if destroyed by dislocation slip) during plastic deformation (to be compared with the much easier glide plane softening discussed later in Section 4.4 for weaker LCOs formed between metallic elements) and as such can provide pinning points for dislocations (Fig. 10a) and divert them to double cross-slip [34,37]. Dislocation movement is rendered sluggish, which facilitates dislocation reactions, multiplication and accumulation, contributing to strain hardening. Indeed, with the OOCs and the randomly distributed individual interstitial solutes, the stress-strain curves show obviously elevated yield strength as well as strain hardening rate and therefore tensile ductility, Fig. 10b [34]. Of course, in this case oxygen embrittlement is a concern, such that the incorporated oxygen content needs to be limited to no more than a couple of at%.

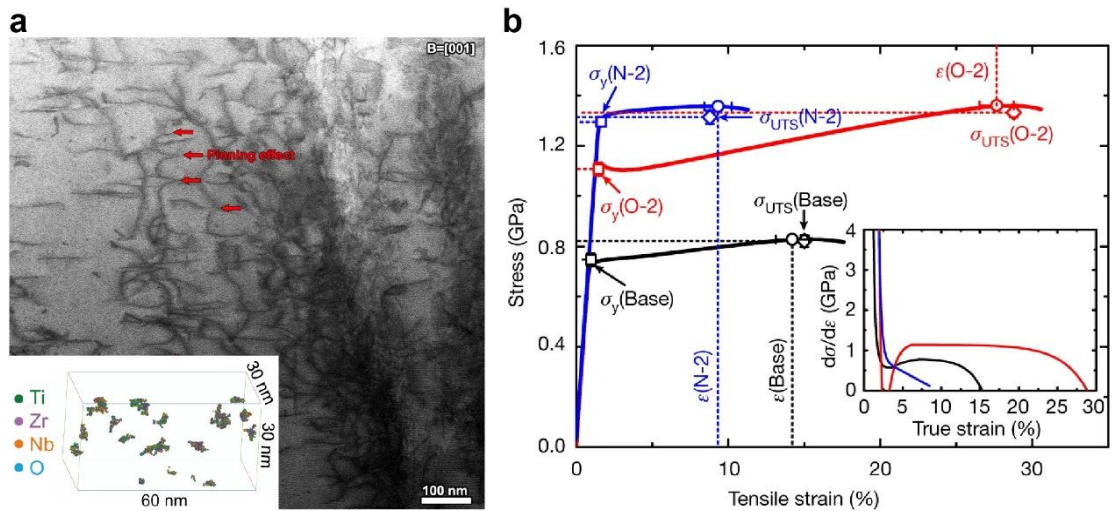

Figure 10. Ordered oxygen complexes (OOCs). (a) STEM observation of the oxygen solutioned alloy after being strained to 8%, indicating dislocation pinning effect via OOCs, adapted from Ref. [34]. The inset is the APT analysis presenting the OOCs enriched in O and Zr/Ti [37]. Compare wtih Fig. S6 with more such LCOs at another composition. (b) Tensile true stress-strain curves of the TiZrHfNb, (TiZrHfNb)$_{98}$O$_2$ (O-2) and (TiZrHfNb)$_{98}$N$_2$ (N-2) alloys, and the inset shows the corresponding strain hardening response [34].

The second type of LCOs are those formed by constituent/substitutional metallic elements, but their effects are not expected to be as pronounced as the LCOs, which involve strong (such as O-metal) bonds. Nonetheless, for LCOs with metallic elements larger ones with dimensions on the scale of several nanometers or more, as well as stripes of variable local composition (and associated SFE) with wavelength on such scale, are expected to be more impactful in interfering with dislocation motion. This is because it is the dimensional scale of such LCO and/or concentration wave that dictates the characteristic length of each dislocation segment stick-slipping forward (e.g., ~ 6 nm in the CoCrFeNiPd HEA [93]), as discussed in Section 4.1. As an example, in an electroplated Ni-Co MPEA with an average composition of $Ni_{50}Co_{50}$ [66], each nanograin contains sub-grain domains of varying composition at a length scale from a few up to 10 nm (as shown in Fig. 3c). This is comparable to the length scale of dislocation segments, which stick-slip through the composition- (and SFE-) modulated regions [58], one at a time (Fig. 9 and Fig. S5b). The slowed-down motion of dislocations promotes their reaction, interlocking and accumulation in the nanometer grains. This produces remarkable strain hardening after yielding, as the density of stored dislocations increases continuously with strain (Fig. S11a, see the TEM image showing the stored dislocations), up till a tensile strength of ~2.3 GPa together with a respectable tensile ductility (Fig. S11b), rivaling and surpassing those of ultra-strong steels. A similar scenario plays out in the spinodal-like undulation in Fig. 3e. There again, the compositional modulation in the bcc solution heightens the ruggedness of the terrain for the passage of dislocations, and hence enhances strain hardening. As a result, the TiVNbHf alloy reveals a high yield strength (1.1 GPa) together with good tensile elongation to failure (28%) [29]. A very recent extension of this idea is shown in Fig. 11. In this case, Al, which has a negative enthalpy of mixing with some of the principal elements such as Hf, has been added. This appears to deepen the two energy minima at particular alloy compositions, such that the spinodal decomposition into two bi-continuous bcc solutions is promoted, one of which is enriched in Hf-Al. In HfNbTiVAl$_{10}$ the authors [94] were able to decrease the modulation to roughly 9 nm wide, and 30 to 100 nm long. A close-up view of such a concentration wave is shown in Fig. 11a. The yield strength (~1.4 GPa) and uniform tensile ductility (~18%) of this HEA are both much elevated relative to those with less Al and less compositional undulation, as compared in Fig. 11b. Note that, having more Al solutes and more relatively strong Hf-Al bonds would be expected to increase yield strength. But here we observe that the strain hardening rate (slope of the curve) is also increased and sustained especially in the later strain stage, thanks to increased degree of chemical inhomogeneity with increasing Al content. The improvements seen in the stress-strain curve/response are now as pronounced as in the case of adding non-metallic interstitial solutes (Fig. 10).

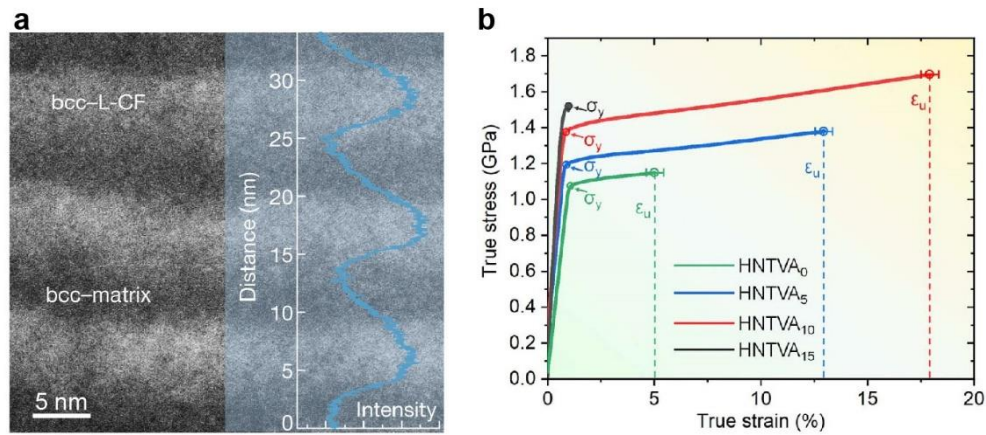

Figure 11. (a) HAADF-STEM image of a bcc HfNbTiVAl$_{10}$ HEA, showing compositional undulation with a wave length around 9 nanometers, which is attributed to spinodal decomposition into bcc-matrix and another bcc phase via long-range chemical fluctuation. (b) True stress–true strain curves comparing the HfNbTiVAl$_0$, HfNbTiVAl$_5$, HfNbTiVAl$_{10}$, and HfNbTiVAl$_{15}$ alloys. Adapted from Ref. [94].

On the high side of the inhomogeneity spectrum, chemical heterogeneity is in the form of fully chemically ordered "precipitates". The LCOs have evolved into "second phase" patches, each a number of nanometers in size, in high volume fraction (almost as populous as the matrix). A prominent example in this regard is given in Ref.

[54], which gives an impressive data point in strength-ductility combination as seen in Fig. 1. In this heterogeneous alloy, dislocations would cut through, rather than bowing out to loop around unshearable particles as in the precipitation hardening case, because the ductile $L1_2$ multicomponent particles are < ~30 nm in size (with ~20 nm gap in between neighboring ones) and perfectly coherent with the fcc matrix. Such a scenario is the "high end" of the picture we have painted above: dislocations have to traverse fairly large (a number of nanometers across) domains of a different composition and chemical order; these "patches" may be perceived as successive speed bump zones built in everywhere on the dislocation pathway.

### 4.4 Delocalization of plastic strain

Finally, when localization of plastic strain does take place at high stresses, chemical inhomogeneity can be made use of to tailor the pattern of the localized shear bands towards improved uniformity in plastic flow. For example, the chemical inhomogeneities can play such a role that their interplay with the evolving lattice distortion in planar slip bands offers a way out towards strain delocalization. This discovery is discussed here using an example in bcc MPEAs, where the presence of LCOs help raise the yield strength to gigapascal level. In particular, how these LCOs affect the stress-strain curve and ductility deserves a careful examination. After yielding, since the LCOs formed between metallic elements have bonding and re-formation tendency weaker than those of OOCs [34,37] discussed earlier, they get gradually destructed by repeated dislocation slips that cut through, which leads to glide plane softening and consequently planar slip that concentrates the strain in a band. One may expect that this would be an outright softening mechanism that eventually causes severe strain localization, which jeopardizes ductility as it is a risky plastic instability. The catch to turn the table around to our advantage is that while LCOs encourage "planar slip band" that localizes the strain, they would also sow seeds to proliferate such bands, which can produce a positive effect on spreading the plastic flow to undeformed sample volume. For this to happen, a mechanism of strain hardening inside the intensifying slip/shear band is imperative (which by the way is unavailable in amorphous alloys such that shear bands there tend to turn into run-away catastrophe [95], due to the lack of any structural strain hardening mechanism). This turns out to be a real possibility in our crystalline lattice, as shown in Fig. 12, via a carefully crafted $Ti_{50}Zr_{18}Nb_{15}V_{12}Al_5$ complex-concentrated composition [36]. This HEA may be viewed as $(Ti_{0.65}Nb_{0.19}V_{0.16})_{77}Zr_{18}Al_5$, a bcc alloy using Nb and V $\beta$-stabilizers, while adding Zr and Al, a pair with a large negative mixing enthalpy that brings a rather strong chemical ordering tendency towards LCO enriched with Zr and Al. In addition to proliferating the B2-like LCO (Fig. 5d), these two species also offer an interesting feature: the larger Zr (0.156 nm) and the smaller Al (0.136 nm) each obviously deviates from the average atomic size of the rest of the elements (0.1435 nm for the base alloy). Before deformation, B2-like LCOs spread out with a spacing of several nanometers (they thus offer ample potential seeding sites to disperse planar slip later). The formation of Zr-Al LCO cancels out the lattice strains caused by each of the two elements, Fig. 12a (see the local region depicted in the middle panel). During deformation, destruction of the LCO due to repeated dislocation slip softens the alloy first. However, the glide plane softening does not continue indefinitely to eventually get out of hand. This is because with continued deformation the LCO dissociates more and more, and the Zr and Al atoms eventually separate to large enough distances. At this point, as shown in Fig. 12a (right panel), the larger Zr produces lattice expansion and the smaller Al renders lattice contraction (Fig. 12a, left panel). The elastic strains from the two species/solutes no longer offset and each contributes separately to hardening (as in traditional solid solution hardening). In other words, the lattice distortion increases in the later stage of plastic deformation — a microstructural mechanism for strain hardening, which combats with the increasing softening due to chemical disordering in the existing band. This balancing act puts a break on outright softening that could run away, as illustrated using a schematic in Fig. S12, opening the door to prevent unharnessed strain localization.

The evolution sequence is shown in Fig. 12b. At a strain of 2.5%, dislocations are constrained in planar slip bands. After the LCO is sufficiently randomized by dislocation, excess lattice strains rise to impose work hardening in these bands. This also helps to store dislocation debris inside the planar slip bands (Fig. S13a), as the population of dislocation loops and tangles dynamically increase. The resultant work hardening encourages dislocations to cross-slip and double-cross-slip away from the first-generation slip bands and form second-generation slip bands (Fig. 12c).

The prolific second-generation slip bands render frequent intersection of the bands with one another throughout the sample, adding even more strain hardening at large strains (Fig. 12d). Consequently, the slip bands proliferate, gradually permeating over 80% of the sample volume. The spread-out plastic flow gives the alloy a ~25% uniform tensile ductility, with ~50% total elongation to failure, both comparable with individual constituent metals inside the HEA. Yet this happens while the multi-component alloy is ~10 times stronger than elemental Nb, see Fig. S13. Note that the above would not be achieved in a sample that starts out with a lower degree of LCO. There, even though the strain hardening inside the first-generation planar slip band would also encourage cross-slip, the cross-slipping dislocations trying to leave the primary bands would tangle into dislocation walls instead, because they have too a low probability to meet the now-less-frequent LCOs to instigate populous second-generation planar slip bands.

To recapitulate, the scenarios discussed in Section 4 underscore an overarching mechanism underlying the strength-ductility synergy. We reiterate that all the landscape-roughening [68] inhomogeneities, even including the fluctuations in a random concentrated solution, not merely elevate strength. Their influence on the evolution of dislocation dynamics can be even more significant than their effect on $\sigma_y$: the plastic flow is rendered increasingly non-homogeneous with large strain gradients [1] as tensile deformation progresses. The inhomogeneities efficiently increase the dynamic microstructure refinement upon deformation with the accumulation of (sometimes geometrically necessary) dislocations [50,51,66] and planar defects, thereby sustaining an adequate $\Theta$. Note that due to the inhomogeneities [65,66,71], all the energy barriers for various activated processes are local properties with a wide distribution in their magnitude. Consequently, different defect mechanisms (such as dislocation propagation, deformation twins, phase transformation) can be activated with comparable propensity at different locations, as the energy barriers (including those that scale with SFE) they encounter locally are often similar in magnitude. Hence, various plasticity (defect) processes get to take place concurrently to generate lots of debris of various kinds at the same time. This keeps adding obstacles for dislocations to pile-up against, increasing long-ranged back stresses that impose non-local hardening. All in all, these roughen the pathway for the running dislocations, raising resistance to their motion while facilitating their reactions and accumulation to result in a high strain hardening capability that sustains large (uniform) tensile elongation at high flow stresses. Meanwhile, the dynamic recovery of dislocations via cross-slip is also reduced [60] due to the low SFE (and hence widened splitting distance between the partials in the extended dislocation), facilitating defect storage to promote strain hardening. Using chemical/structural heterogeneities to provide extensive work hardening thus offers an effective strategy to strengthen without much sacrifice in ductility. The summary of strength-ductility for the recent HEAs shows an upside-down banana curve at least for yield strength up to 1.3 GPa (see Fig. 1), which is pushing forward the strength-ductility synergy and furthers the course chartered by previous high-performance alloys.

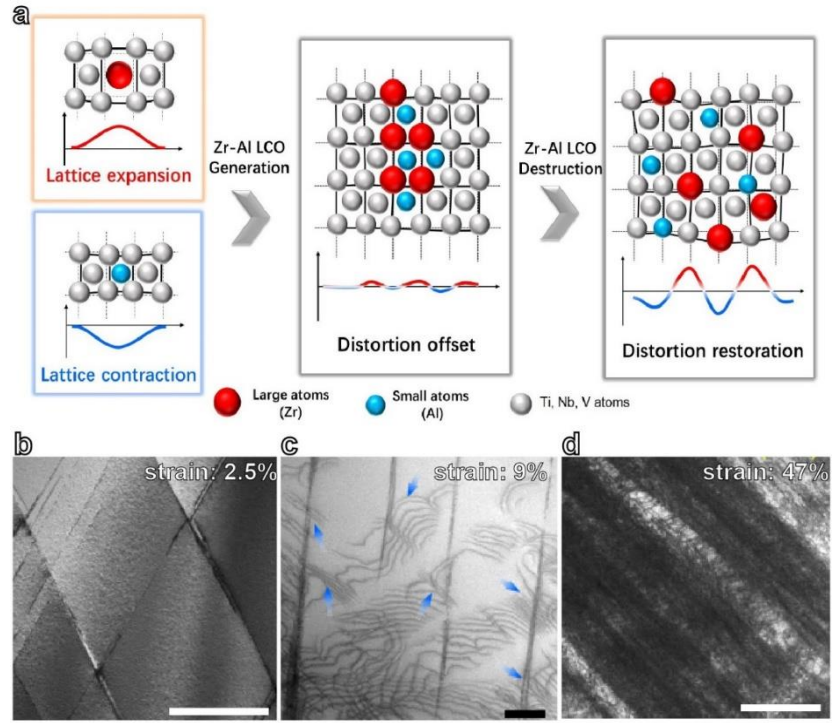

Figure 12. Plastic deformation mechanism in a bcc $Ti_{50}Zr_{18}Nb_{15}V_{12}Al_5$ (at%) MPEA [36]. (a) Schematic diagram demonstrating that the formation of LCO comprising Zr and Al offsets the lattice distortion, and the deformation-induced LCO destruction promotes distortion through randomizing and separating Zr and Al atoms. (b-d) STEM images of the alloy at various strains. (b) Planar-slip bands are formed at the early stage of straining (2.5% strain). (c) Dislocations glide away from the first-generation bands and form second-generation bands as the strain was increased to 9%. The blue arrows highlight the dislocation pile-ups adjacent to the first-generation bands. (d) Permeating slip bands were formed throughout the sample, spreading plastic flow to over 80% of the volume.

## 5. Concluding remarks and outlook

This review aims to deliver the following take-home message. The previous practice/successes using heterogeneous microstructures and high-content of alloying (including interstitial) elements have not yet exhausted the room for improvement, in terms of mitigating the trade-off between the yield strength and tensile ductility. We can now solicit additional help by resorting to "high-entropy alloys". These MPEAs aid in producing, and embedding into the microstructure, an unusually rich variety of spatially closely spaced heterogeneities [60] that increase the resistance to dislocation motion and the propensity for defect accumulation, on the fly during tensile deformation. To hammer home our message, we recapitulate that the exceptionally concentrated make-up of a HEA inherently imparts "nano-cocktail"-like compositional inhomogeneity on the scale of one nanometer in the random solution case, and LCOs from sub-nanometer to a few nanometers, not to mention stripes/patches on the scale of a number of nanometers in the cases where undulation/LCOs are artificially installed inside the alloy microstructure. In all these cases, the nanoscale segment detrapping (NSD) mechanism for dislocation motion ushers in major differences from previous metals and conventional alloys via unusual "nano-cocktail hardening" [82]. The NSD not just elevates the strength due to bumpy dislocation ride across a chemically-sensitive and complex landscape, but more importantly, it features a new form of stick-slip conducive to dislocation stall, facilitating tangling and accumulation. Generally speaking, when additional inhomogeneities are intentionally introduced, in particular tailorable chemical varieties in terms of composition undulation and local chemical order, beyond the structural ones such as a distribution of grain size and phases known before, (more and harder) dislocation obstacles/traps are dynamically embedded into an increasingly rugged landscape, to keep strain hardening going long after yielding. The work hardening rate $\Theta$, although perhaps not always at a level as high as that in an un-strengthened simple fcc metal, becomes nevertheless sufficient to keep up with the high and rising flow stress, often via multiple (twinning/transformation/dislocation accumulation) strain-hardening stages [54]. Comparatively speaking, relative to elevating the initial yield strength, inhomogeneity engineering is arguably more effective/powerful in changing the shape of the stress-strain curve. This is done by regulating the evolution of dislocations (and hence strain hardening) to help spread the strain throughout the material undergoing plastic deformation. In this regard, the innovation emphasized in this review is the intelligent design to deploy additional chemical inhomogeneities.

In terms of general materials science insight, a new feature ushered in by concentrated MPEAs due to chemical inhomogeneities, is the wide variability (spread and distribution) of the energy barrier for each and every activated (dislocation) process. These barrier distributions overlap, reshaping the balance/competition between various dislocation-obstacle interactions. The synergistic effects on defect motion (strengthening) and accumulation (strain hardening) in this review are just one example.

As seen in Fig. 1, many HEAs offer sufficient strain hardening that prolongs tensile ductility, such that coexisting high (GPa) yield strength and pure-metal-like (un-strengthened coarse-grained elemental) ductility, previously conjectured to be mutually exclusive, is largely resolved. This circumvents the strength-ductility trade-off across a considerable range of strengthening at least to 1-GPa level, up to which the $\sigma_y$ can be reached without overly painstaking efforts. The mechanical properties of some non-ferrous alloys can now outrank state-of-the-art steels. The new envelopes in the property map (Fig. 1), pushed greatly to expand the materials selection repertoire, offer new opportunities to facilitate engineering design.

To put things in perspective, we are aware that in the literature there were many reports about overcoming the strength–ductility trade-off. But the vast majority of those successes have been claimed with respect to a microstructural state after considerable hardening treatment [10,34,59], which has already seriously compromised ductility. In other words, their starting reference is far away from a coarse-grained metal but resides around the elbow of the banana-shaped curve (Fig. 1), such that a lift out of that region towards the upper-right direction (e.g., the rule-of-mixtures interpolation in Fig. S2a), could already be deemed as successfully evading the trade-off. While this has helped in arriving at a better strength-ductility combination, the mission is far from being fully accomplished, as illustrated in Fig. 1. With HEAs, now we have reached a different and higher level: the strengthening and strain hardening mechanisms afforded by solid solutions have been further optimized, and the properties can hence be pushed to the point that $\sigma_y$ is raised by many folds while the ductility remains as high as that of an un-strengthened metal, or to the point of a $\sigma_y$ above the GPa benchmark, and yet the ductility can be maintained to approach that (of the order of 50%) of elemental metals. This is therefore a major step-forward towards truly resolving the strength–ductility paradox.

As an outlook, the next challenge is to use the new insight (the design/manipulation of inhomogeneities) to evade (or at least alleviate to a greater extent) the trade-off that currently persists for super-high $\sigma_y$ (the high end of the spectrum is of the order of >2 GPa [74,96]) alloys. This is a tall order and may require yet another level of innovative microstructural design to build the hierarchy of inhomogeneities and devise mechanisms for them to persist or even dynamically self-reinforce during tensile deformation. In terms of structural inhomogeneities, one can start with multimodal grain sizes [7] and segregate certain species to grain boundaries to stabilize grain size distribution. One can also change constituent elements to increase lattice distortion (an atomic-scale inhomogeneity). With regard to how to chemically enhance inhomogeneities, several routes have been outlined in this review. For example, alloying elements can be judiciously selected to lower the SFE, which promotes planar defects such as SFs and nanotwins. Compositional undulations can be intentionally designed to introduce chemical "waves", with amplitude and length scale well beyond statistical fluctuation. In particular, the wave length can reach many nanometers via spinodally decomposed phases [94] or artificially modulated deposition [66]. Domains of LCOs up to nanometer-scale patches can be created as well. Note that these chemical undulations are more populous when distinctly different substitutional/interstitial solutes are introduced into the HEA solution, because a large disparity in chemical affinity (mixing enthalpy [94]) between the added element and the constituent species helps to intensify local chemical order and favor certain compositions.

Along this line, the recently discovered dislocation cells, which are almost a ubiquitous feature in 3D-printed (particularly fcc) alloys [97], seem to add a new knob to turn. The dislocation cellular structures are sometimes also found in heavily deformed metals [44,98]. It would be interesting to engineer the dimensions of these cells, going from the hundreds of nanometers at present down to a few tens of nanometers, so that they contribute appreciably to strengthening. Meanwhile, some sessile dislocations at the cell boundaries can be expected to evolve into reinforced dislocation walls along with increasing tensile straining, adding a vehicle to improve strain hardening. Here we note again that our goal is not to reach a strain hardening rate as high as that initially available in an un-deformed coarse-grained metal, but rather a sufficient $\theta$ sustained to delay strain localization stability at very high flow stress level. All in all, we project continued endeavors in pushing forward the frontier of science with regard to the strength-ductility synergy, in keeping with human beings' never-ending quest to strive for higher and better.

Last but not least, the strength-ductility synergy via the unusual strengthening/dislocation slip mechanisms discussed in this review is made possible by concentrated complex alloys (CCAs) exploiting multiple principal elements. This success further incentivizes the ongoing HEA research. After all, it is the pursuit for unprecedented mechanisms/properties in hitherto-unexplored composition space that has been the primary motivation driving forward the HEA field. In retrospect, some of the excellent properties reported for HEAs in recent years, such as the high fracture toughness and coexisting high strength and high ductility of fcc MPEAs at cryogenic temperatures, are not all that exceptional as claimed when compared with previous best-performing fcc metals and solid solution alloys such as 316LN stainless steel [57] and MP35 (Fe-35Co-35Ni-20Cr-10Mo, wt%) [99] cryogenic alloys. In this regard, this review serves the purpose of illustrating how MPEAs, leaning towards an extreme of concentrated solutions, push forward the multi-stage strain hardening mechanisms including the known TWIP and TRIP to yet another level (by tuning the SFE or the metastability of the matrix phase to promote twinning or martensitic transformation). Our specific example in this review is the unusual "nano-cocktail" (chemical inhomogeneities from CSRO and nanometer-wavelength undulations all the way to unusually high content of interfaces) solution hardening mechanism accompanied by an increased strain hardening capability, leading to unprecedented strength-ductility synergy.

**Acknowledgement**

The support from National Natural Science Foundation of China (Grant No. 52231001) is gratefully acknowledged. C.L. acknowledges support from National Natural Science Foundation of China (Grant No. 52371162) and National Natural Science Fund for Excellent Young Scientists Fund Program (Overseas). E.M. and C.L. acknowledge XJTU for hosting their research at the Center for Alloy Innovation and Design (CAID).

**Supplementary Information**

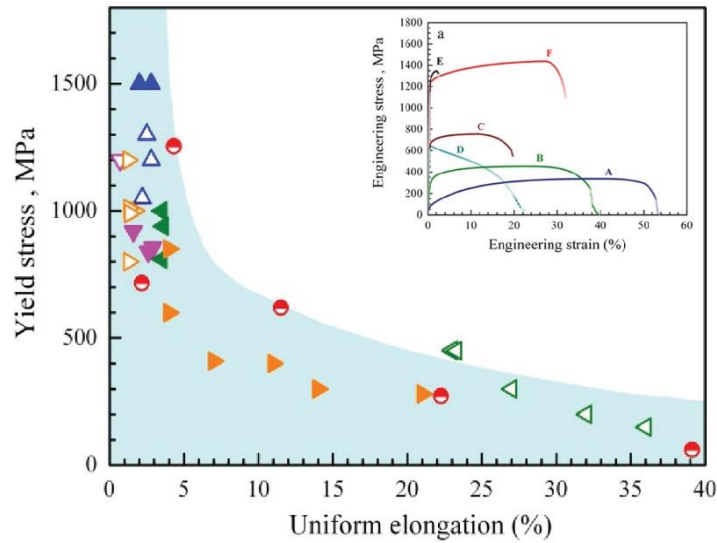

Figure S1. Yield strength and uniform tensile elongation for Ni metal processed to various microstructures, adapted from a number of published reports in the literature. The top-right inset presents representative tensile engineering stress-strain curves of Ni, with different grain sizes after microstructural refinement via e.g., cold rolling or severe plastic deformation, including coarse-grained (27 μm in grain size, curve A), ultrafine grained (200 nm, curve C), nanocrystalline grained (18 nm, curve E), and nanodomained (7 nm-diameter, curve F) samples [S1]. From case A to case E, the strengthening is obvious, showing rising yield strength, whereas the tensile elongation to failure decreases. Note that this apparent strength-ductility trade-off appears while the dislocations continue to function as viable plasticity carriers: they remain widely active and are responsible for the appreciable tensile strains achieved (see, e.g., curve B and curve C).

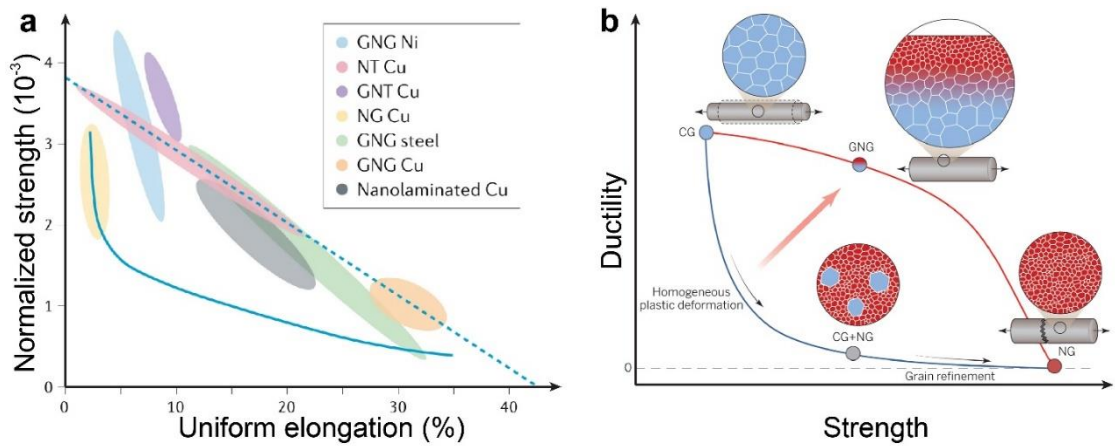

Figure S2. Strength-ductility space for metals and alloys (adapted from [S2] and [S3]). (a) Normalized yield strength versus uniform tensile elongation for gradient nanostructured and homogeneous metals and alloys. The strength is referenced relative to that of the coarse-grained counterpart, adapted from [S2]. The improved properties due to the heterogeneous/gradient nanostructure fall close to the dashed line representing the rule of mixtures. (b) Schematic showing strength-ductility synergy, adapted from [S3]. The red line sketched there, while not yet realized in experiments summarized in (a), represents a combination of nanograin-like strength and coarse-grain-like ductility, which has long been the target of material scientists.

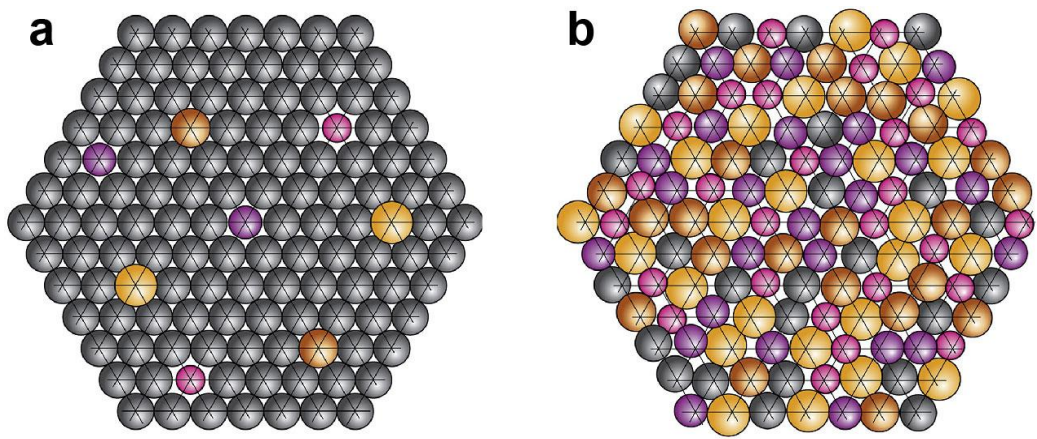

Fig. S3 Atom positions in (a) a dilute solution and (b) a concentrated solution, adapted from [S4]. In the dilute solution (a), the atoms are constrained to occupy lattice sites by surrounding solvent atoms; while in the concentrated solution (b), the atoms often deviate from the lattice positions.

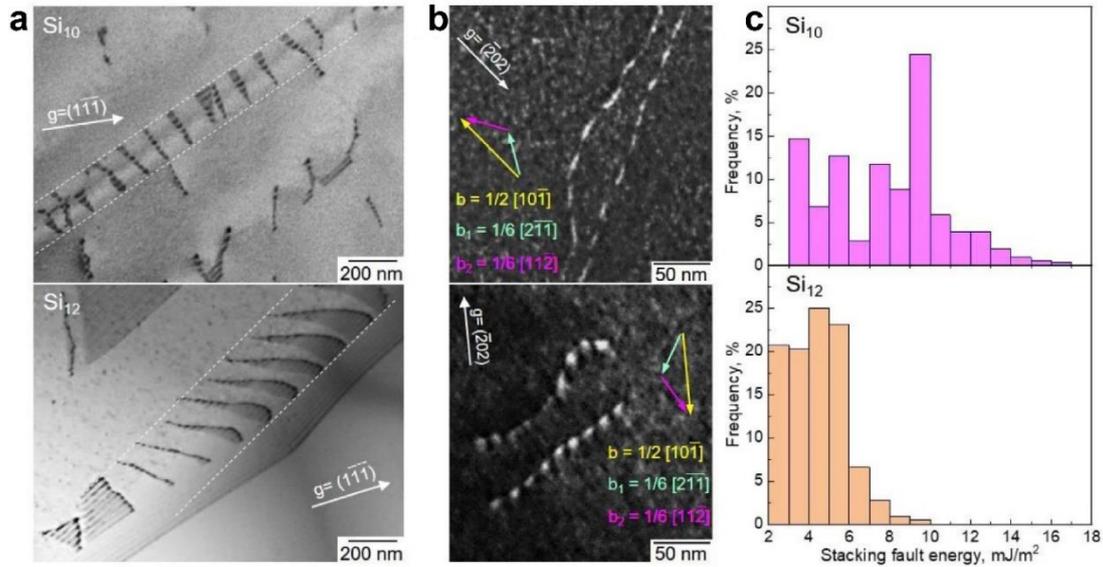

Figure S4. Stacking fault analysis of the Si-10 ($Co_{20}Cr_{20}Fe_{20}Ni_{20}Mn_{10}Si_{10}$) and Si-12 ($Co_{22}Cr_{22}Fe_{22}Ni_{22}Si_{12}$) HEAs [S5]. (a) Two-beam bright-field transmission electron microscopy (TEM) images presenting the accumulation of dislocations. (b) Two-beam dark-field TEM image showing pairs of the $1/6<112>\{111\}$-type partial dislocations, showing large and variable split distance. (c) The stacking fault energy (SFE) was estimated to be mostly below 10 mJ/m$^2$, from the various dislocation dissociation widths measured for the extended dislocation at different locations under TEM. Apparently, the chemical tuning here (unusually large amounts of substituting Si added into the HEA) lowers the SFE to single digit. The wavy dislocation line hints at different local chemical composition and local SFE.

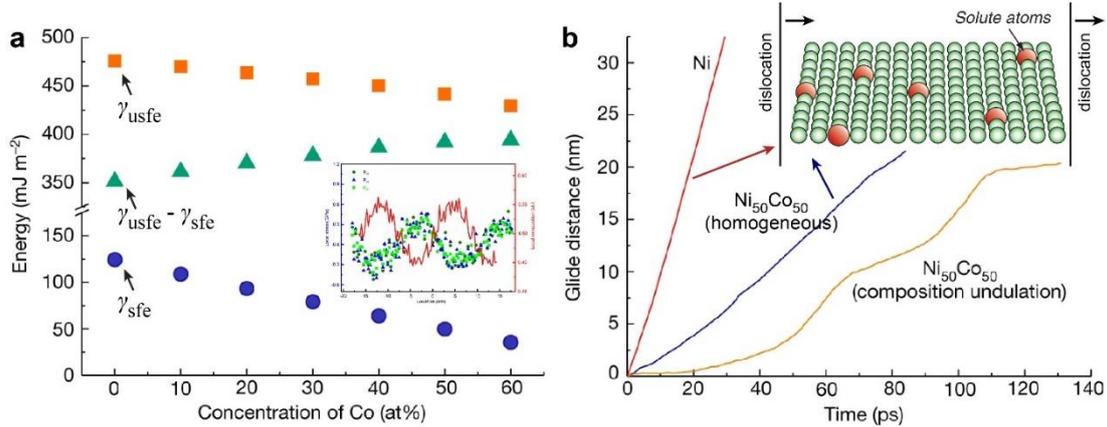

Figure S5. Influence of compositional undulation on dislocation dynamics in concentrated Ni-Co solid solutions [S6]. (a) SFE (circles), unstable SFE (squares) and their difference (triangles) as a function of Co concentration. The inset in (a) presents the composition undulation with a wavelength of 10 nm (red curve), which is highly correlated with the local stress (dots) in the Ni-Co alloy. Such a purposely electroplated composition undulation in Ni-Co (concentrated, and "random" on average) fcc solid solution significantly widens the spatially variable local SFE; for example, at $Ni_{60}Co_{40}$ the local SFE is about twice that at $Ni_{40}Co_{60}$ (in contrast, for a Ni-based solid solution with only a few at% Co solutes dissolved in it, the SFE would be by and large uniform). (b) Molecular-dynamics-simulated dependence of glide distance versus time for a <110>/2 dissociated edge dislocation under a constant shear stress, contrasting elemental Ni and homogenous $Ni_{50}Co_{50}$, both having a constant speed (slope in this plot), with the "stop-and-go" behavior in compositional undulated $Ni_{50}Co_{50}$ solution. The top-right inset in (b) is a schematic showing the traditional dilute solid solution hardening mechanism, where a dislocation line moves smoothly across a homogeneous and dilute solid solution and remains straight after interactions with the local elastic strain field (due to misfit with the host solvent) of individual solute atoms one at a time. As depicted in Fig. S5b, the motion of dislocations in inhomogeneous concentrated solutions is far more intermittent than in homogeneous solid solutions and normal metals. We see clear "stop-and-go", which is what we mean by stick slip (yellow curve). This is different from the standard dislocation interaction with individual solute atom in a dilute solid solution, or interaction with a truly random solution without LCO and chemical fluctuation. In the latter cases, the drag on the moving dislocation is much weaker and the "hold" is of a much shorter duration, such that the distance vs time plot remains linear, giving an almost constant dislocation speed (blue curve).

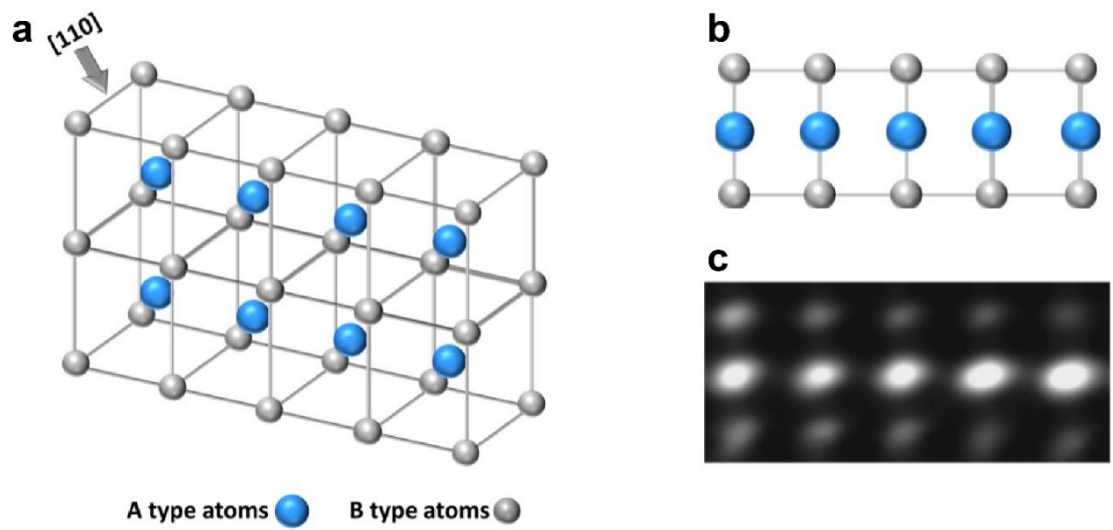

Figure S6. Comparison of B2 lattice and the B2-like LCO in the bcc $Ti_{50}Zr_{18}Nb_{15}V_{12}Al_5$ multicomponent alloy [S7]. (a) A schematic of the B2 lattice. (b) Projection of the B2 lattice along the [110] direction. (c) HAADF-STEM image of the $Ti_{50}Zr_{18}Nb_{15}V_{12}Al_5$ alloy. Due to B2 chemical order involving mostly Zr and Al, the Z contrast intensities of the middle atom-columns (Zr-enriched) are obviously higher than those in the (Al-enriched) planes above and below. The TEM-observed pattern in (c) is clearly consistent with the B2 projection in (b).

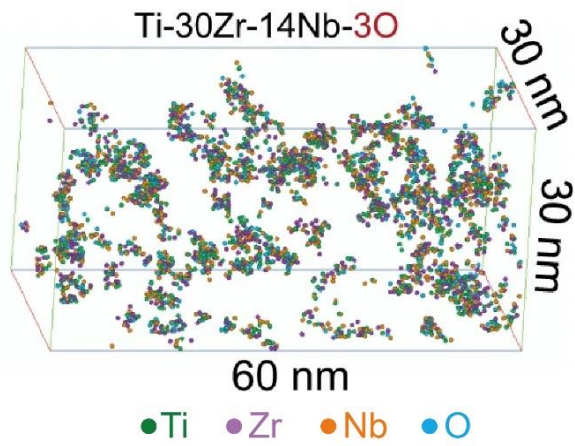

Figure S7. Populous nanometer-scale local ordered-oxygen-complexes (OOCs) in a Ti-30Zr-14Nb-3O (at%) alloy [S8]. The high chemical affinity of oxygen with Zr and Ti induces local chemical inhomogeneity in terms of O-Zr (Ti) enriched LCO and accompanying local compositional fluctuation.

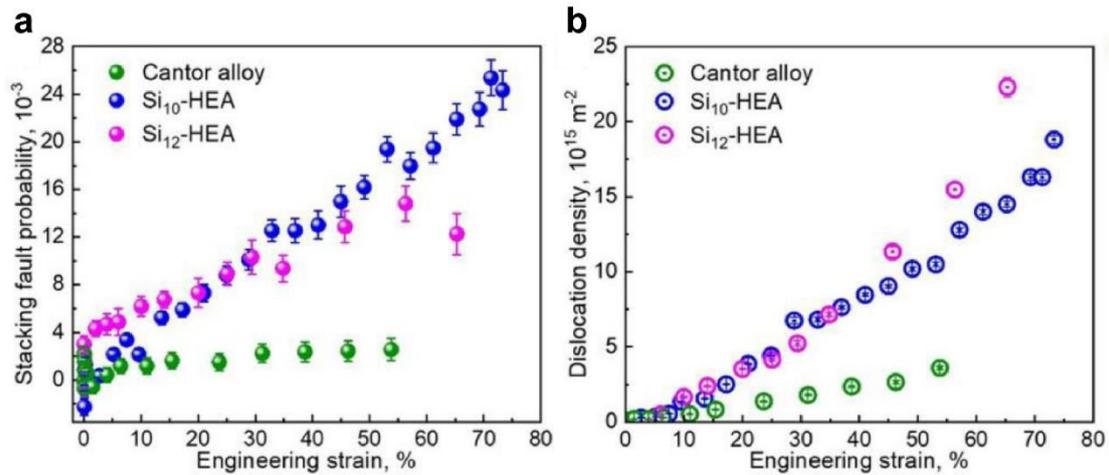

Figure S8. Defect accumulation in the Cantor alloy (which is reminiscent of that in austenitic stainless steels such as 316LN), Si-10 ($Co_{20}Cr_{20}Fe_{20}Ni_{20}Mn_{10}Si_{10}$) and Si-12 ($Co_{22}Cr_{22}Fe_{22}Ni_{22}Si_{12}$) HEAs in tensile deformation, measured using in-situ neutron diffraction [S5]. (a) Stacking fault probability and (b) Dislocation density as a function of plastic strains, showing much elevated defect storage during tensile straining in the chemically modified HEAs relative to the base Cantor HEA. This embeds boundaries/obstacles on the fly during tensile straining, sub-dividing the microstructure and making it increasingly inhomogeneous to promote strain hardening.

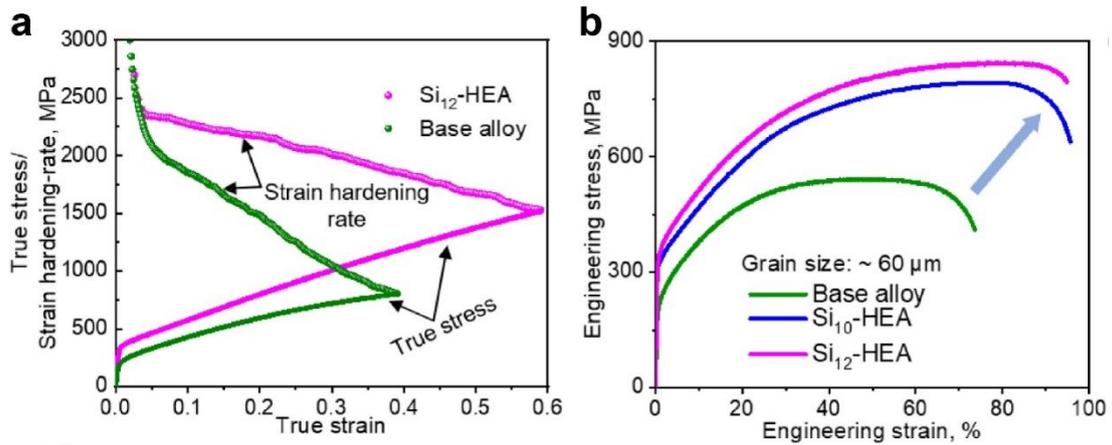

Figure S9. Mechanical properties of the Cantor alloy, Si-10 and Si-12 HEAs [S5]. (a) True stress-strain curve and strain hardening rates during tensile tests. The high defect storage rate shown in Fig. S9 keeps up the strain hardening rate to large strains, much more so than in the base Cantor alloy. (b) Engineering stress-strain curves in uniaxial tensile test. There is no trade-off to worry about: the strength and ductility synergistically rise at the same time.

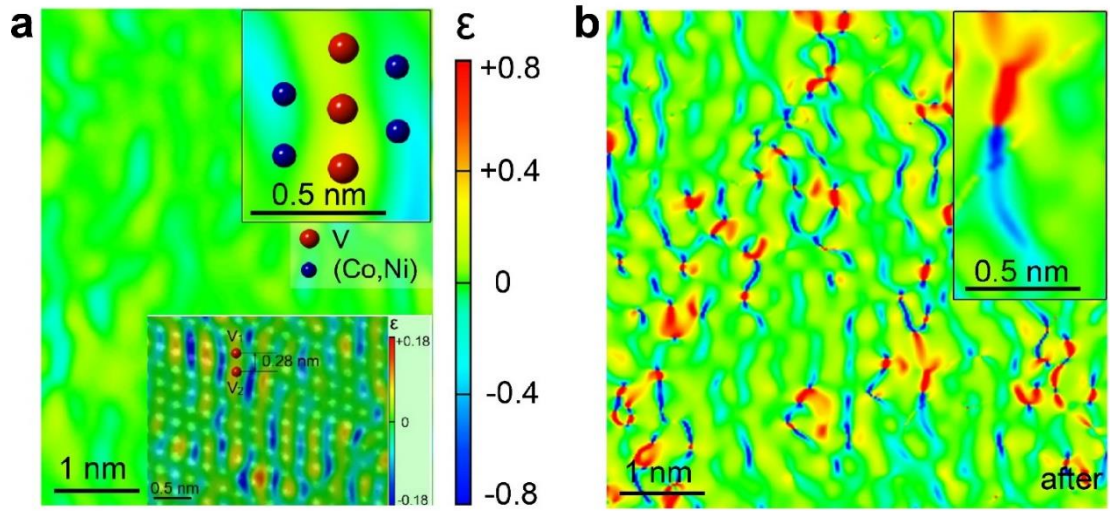

Figure S10. Strain mapping of the VCoNi MEA [S9]. (a) Chemical short-range orders (CSROs) with associated elastic strains. The inset in (a) indicates that the positive strain (yellow band) is caused by the V-enriched column in the CSRO. (b) Strain mapping of the sample after tensile testing. The inset in (b) is a close-up view showing the dislocation-induced strain field (red-blue), which overlap with the strains (yellow) due to CSRO. This suggests that the CSROs could help retard the motion of dislocations and promote their accumulation in the lattice during deformation.

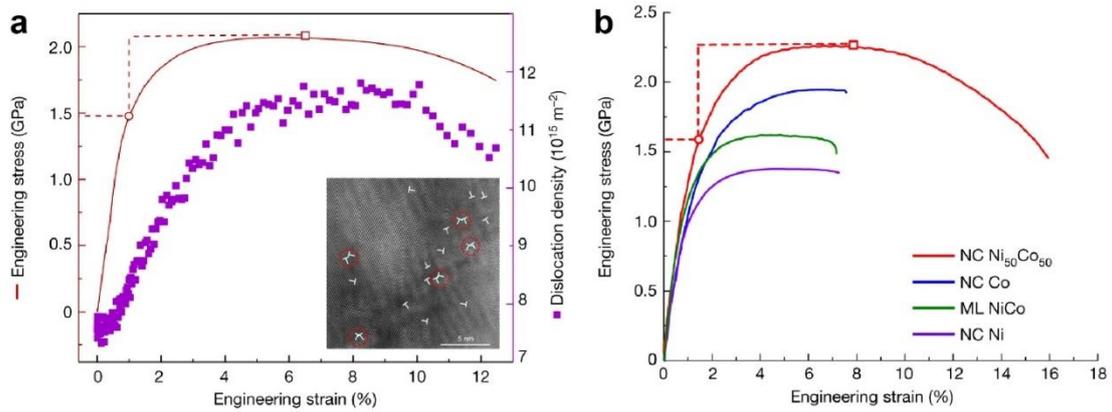

Figure S11. Mechanical properties of the electroplated NiCo alloy with composition undulation on the lengthscale of one to 10 nanometers [S6]. (a) Engineering stress-strain curve and the corresponding dislocation density (purple-color) as a function of strain in tension. The inset is the inverse FFT image showing that a high density of dislocations has been effectively stored in the nanograins, with the red circles highlighting the profuse formation of Lomer locks. (b) Engineering stress-strain curves, showing the extraordinarily high strength and decent ductility of the NiCo alloy. This hierarchically nanostructured alloy, while just a single-phase fcc solid solution, rivals ultra-strong steels in both tensile strength and ductility.

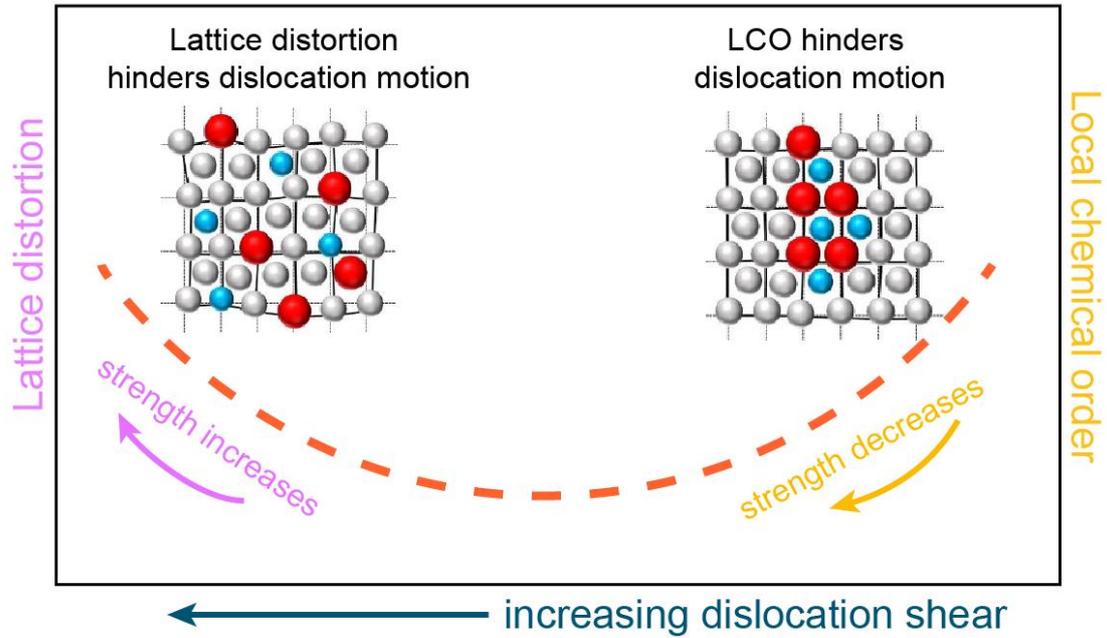

Figure S12. Hardening in the planar slip band due to lattice distortion strains induced by the individual solutes (see Fig. 12 and their atomic sizes in the main text), arising from dissociating Zr (red balls)-Al (light blue balls) LCO during plastic deformation (decreased LCO due to repeated dislocation shear). This puts a break on outright softening due to the destruction of LCOs in the band, and together with other strain hardening processes discussed in the main text pre-empts run-away strain localization.

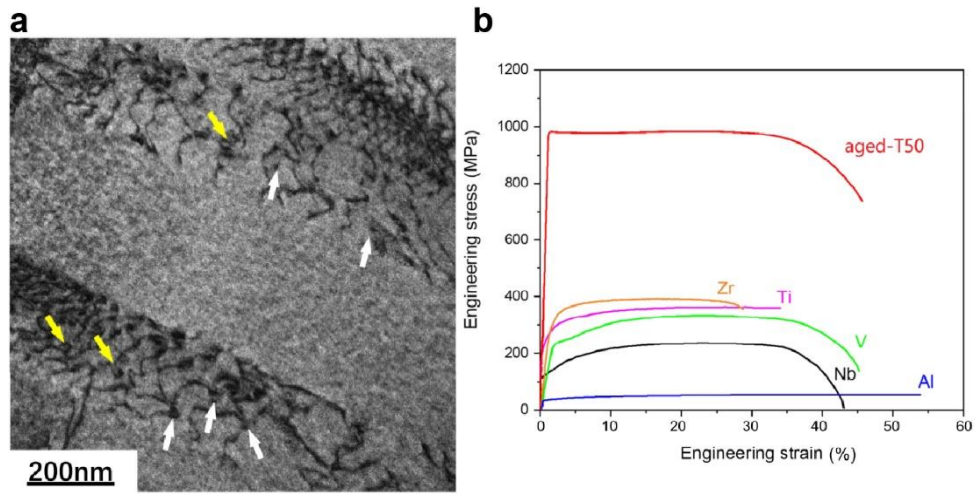

Figure S13. Dislocation behavior and tensile properties of the bcc $Ti_{50}Zr_{18}Nb_{15}V_{12}Al_5$ multicomponent alloy, adapted from [S7]. (a) STEM image of the aged HEA upon tensile deformation to 5% strain. Arrows highlight dislocation debris (yellow) and dislocation loops (white) inside the planar slip bands. (b) Engineering stress-strain curves of the aged- $Ti_{50}Zr_{18}Nb_{15}V_{12}Al_5$ alloy, in comparison with those of elemental metals (Ti, Zr, Nb, V, and Al). The aged $Ti_{50}Zr_{18}Nb_{15}V_{12}Al_5$ alloy retains a ductility similar to that of the constituent elemental metals, while the strength (at GPa level) is about ten times higher, when compared with, e.g., Nb.